\documentclass[twocolumn,aps,pra,nobibnotes,10pt]{revtex4-1}
\usepackage{amsmath}
\usepackage{bbold}
\usepackage{epsf}
\usepackage{graphicx}
\usepackage{xcolor}

\begin{document}

\title{Exchange interaction, disorder, and stacking faults in rhombohedral graphene multilayers}

\author{James H. Muten}
\affiliation{Physics Department, Lancaster University, Lancaster, LA1 4YB, UK}

\author{Alex J. Copeland}
\affiliation{Physics Department, Lancaster University, Lancaster, LA1 4YB, UK}

\author{Edward McCann}
\email{ed.mccann@lancaster.ac.uk}
\affiliation{Physics Department, Lancaster University, Lancaster, LA1 4YB, UK}

\begin{abstract}
We apply the mean-field Hartree Fock theory of gapped electronic states at charge neutrality in bilayer graphene to thin films of rhombohedral graphite with up to thirty layers. For the ground state, the order parameter (the separation of bands at the valley center) saturates to a constant non-zero value as the layer number increases, whereas the band gap decreases with layer number. We consider chiral symmetry breaking disorder in the form of random layer potentials and chiral preserving disorder in the form of random values of the interlayer coupling. The former reduces the magnitude of the mean band gap whereas the latter has a negligible effect, which is due to self-averaging within a film with a large number of layers. We determine the ground state in the presence of an individual stacking fault which results in two pairs of low-energy bands and we identify two separate order parameters. One of them determines the band gap at zero temperature, the other determines the critical temperature leading, overall, to a temperature dependence of the band gap that is distinct to that of pristine rhombohedral graphite. In the presence of stacking faults, each individual rhombohedral section with $m$ layers contributes a pair of low-energy flat bands producing a peak in the Berry curvature located at a characteristic $m$-dependent wave vector. The Chern number per spin-valley flavor for the filled valence bands in the ground state is equal in magnitude to the total number of layers divided by two, the same value as for pristine rhombohedral graphite.
\end{abstract}

\maketitle

\section{Introduction}

\begin{figure}[t]
\includegraphics[scale=0.32]{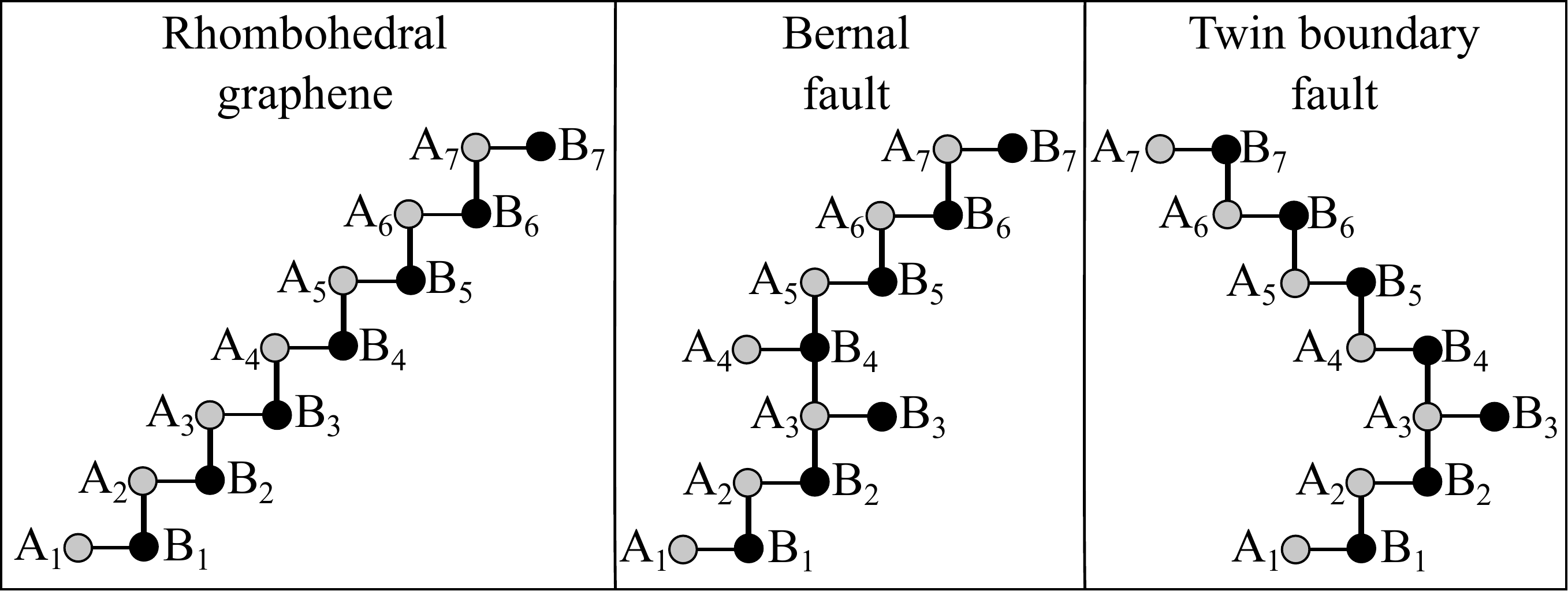}
\caption{Schematic side view of the lattice of rhombohedral graphene with $N= 7$ layers showing the pristine lattice, a lattice with a single Bernal stacking fault, and a lattice with a single twin boundary stacking fault. Labels indicate the $A_n$, $B_n$ atomic sites on the $n$th layer, horizontal solid lines indicate intralayer hopping with parameter $\gamma_0$, vertical solid lines indicate interlayer hopping $\gamma_1$.
}\label{lattice}
\end{figure}

Recently, topological flat bands have been the subject of intense research in twisted bilayer graphene~\cite{cao18a,cao18b} as well as other two-dimensional systems including the Lieb, honeycomb and kagome lattices~\cite{sutherland86,lieb89,wu07,guo09,sun11,drost17,leykam18,kempkes19,kang20}. There are also topological flat bands in rhombohedral multilayer graphene (RMG)~\cite{mcclure69,latil06,aoki07,arovas08,koshino09,zhang11,slizovskiy19} in which alternating intra- and interlayer coupling act like staggered hopping in the Su-Schrieffer-Heeger (SSH) model~\cite{su79,asboth16,heikkila11,xiao11}, as shown in Fig.~\ref{lattice}.
Now there is fresh interest in RMG due to progress in fabricating and characterizing samples with a large layer number~\cite{pierucci15,henni16,henck18,lat19,yang19,geisenhof19,lee19,bouhafs20,kerelsky21} culminating in the realization of high-quality films with up to fifty layers~\cite{shi20}.

In high mobility samples, at charge neutrality, low temperature and for zero external fields, low energy bands have been observed to be gapped in bilayer graphene~\cite{weitz10,freitag12,velasco12,bao12,veligura12,freitag13}, Bernal multilayers with up to $N=8$ layers~\cite{grushina15,nam16,nam18}, rhombohedral multilayers with up to $N=4$ layers~\cite{bao11,lee14,myhro18} and recently with $N \approx 12$~\cite{shi20}. A number of different interaction-induced broken symmetry states have been proposed for bilayer graphene~\cite{nilsson06,min08,sun09,vafek-yang10,nandkishore10,zhang10b,nandkishore10b,vafek10,lemonik10,zhang11,jung11,macdonald12,lemonik12,kharitonov12,cvetkovic12,zhu13} including pseudospin layer antiferromagnetic (AF) states~\cite{min08,zhang10b,zhang11,jung11,macdonald12} in which electrons with different spin and valley flavors spontaneously accumulate on different layers, creating an odd parity state that breaks inversion symmetry and opens a gap. Owing to the antiferromagnetic configuration of four flavors, there is no net charge accumulation on summing over them and, thus, no cost in terms of Hartree energy.
The evolution of similar gapped states with layer number has been discussed for both Bernal~\cite{yoon17,koshino17} and rhombohedral multilayers~\cite{zhang11,jung13,jia13,pamuk17}.

In this paper, we apply the Hartree Fock mean-field theory~\cite{min08,jung11,jung13,yoon17,koshino17} of the pseudospin AF state to RMG with a large number of layers (up to $N=30$) in order to determine the layer dependence of the interaction-induced band gap at charge neutrality. We find that the strength of the interacting state, as characterized by the separation of the bands exactly at the valley center, saturates with layer number, but that the actual band gap decreases with layer number; this agrees with density functional theory (DFT) which predicted the band gap for RMG with up to eight layers~\cite{pamuk17}.

We then consider the robustness of the AF state to defects by including values of tight-binding parameters that are constant within each layer (thus preserving translational invariance within each layer) but that vary randomly between layers. We compare disorder that preserves chiral symmetry (random values of the interlayer hoppings) with disorder that breaks chiral symmetry (random layer potentials). For weak disorder, we find that the mean band gap is diminished by chiral breaking disorder, but it is almost insensitive to chiral preserving disorder; this is similar to the behavior of gapless edge states in the non-interacting SSH model~\cite{mondragonshem14,liu18,perezgonzalez19,jurss19,scollon20}.

Another type of defect is a localized stacking fault~\cite{taut14,garciaruiz19,shi20} within a large RMG system, namely a Bernal fault or a twin boundary fault, Fig.~\ref{lattice}. They are particularly interesting because they introduce additional flat bands into the energy spectrum. For a Bernal fault, we find that it introduces a weak connection between two sections of RMG and that the interacting ground state is a straightforward generalization of the AF state with odd parity. The twin boundary fault, however, creates a stronger connection between two RMG sections: the interacting ground state is also a AF state, but with even parity within each spin-valley flavor. For both of these types of ground state, the Chern number per spin-valley flavor has magnitude $N/2$, as is the case for pristine RMG~\cite{fukui05,asboth16,chernvalley}.

With two pairs of flat bands near low energy in a system with a single stacking fault, we identify two order parameters: $\Delta_1$ is the separation at $k=0$ of the lowest conduction band and the highest valence band and $\Delta_2$ is the separation at $k=0$ of the second lowest conduction band and the second highest valence band. For a stacking fault that splits RMG into two sections, $\Delta_1$ is attributed to the shorter section, $\Delta_2$ to the longer one. Although $\Delta_1 \leq \Delta_2$, we find that the transition temperature for the AF state is determined by $\Delta_2$. The temperature dependence of $\Delta_2$ resembles that of an isolated section of RMG, whereas the temperature dependence of $\Delta_1$ is affected by proximity to the longer section. This ensures that $\Delta_1$ (and the overall band gap) remains non-zero up to the relatively high $T_c$ determined by the longer section and $\Delta_2$, and the temperature dependence of $\Delta_1$ (and the band gap) is generally quite distinct from that of pristine RMG.

Section~\ref{method} describes the methodology including the non-interacting Hamiltonian and the Hartree Fock mean-field theory. We use the minimal model with nearest-neighbor intralayer and interlayer hopping parameterized by $\gamma_0$ and $\gamma_1$, respectively, but neglecting other tight-binding parameters. This is done for simplicity and, in particular, it dramatically simplifies the calculation of the exchange interaction allowing us to consider large layer number $N \gg 1$.
Section~\ref{pristine} describes the AF state in pristine RMG. We introduce a toy two-band model that can be solved analytically to give simple expressions for the parameter dependence of the band gap that are broadly in qualitative agreement with the full numerical model. With the full numerical model, we determine the layer and temperature dependence of the band gap. Then, our main results are described in Section~\ref{disorder} for disorder, Section~\ref{bernal} for the Bernal stacking fault, and Section~\ref{twin} for the twin boundary fault.
Finally, in Section~\ref{temp}, we determine the temperature dependence of the order parameters of the AF state for a single stacking fault, Bernal or twin boundary.

\section{Methodology}\label{method}

\subsection{Effective mass model}

The lattice of RMG with $N$ layers consists of two inequivalent sites $A_n$, $B_n$ on each layer, $n = 1, 2, \ldots N$, with sites $B_n$ located below $A_{n+1}$, Fig.~\ref{lattice}. In the tight-binding model, interlayer coupling between $p_z$ orbitals on the $B_n$ and $A_{n+1}$ sites hybridizes those orbitals leading to gapped bulk conduction and valence bands. In the surface layers, however, the $A_1$ and $B_N$ sites don't have neighbors in the next layers so their $p_z$ orbitals aren't hybridized by interlayer coupling, resulting in low-energy surface states within the bulk gap in the vicinity of each of two valleys $K_{\pm 1}$.

In a basis of $p_z$ orbitals on $A_1$, $B_1$, $A_2$, $B_2$, \ldots , $A_N$, $B_N$ sites, the non-interacting Hamiltonian of RMG with $N$ layers ~\cite{mcclure69,arovas08,koshino09} may be written near each valley as
\begin{eqnarray}
H = 
\begin{pmatrix}
 D & V & 0 & 0 & \cdots \\
 V^{\dagger} & D & V & 0 & \cdots \\
 0 & V^\dagger & D & V & \cdots \\
0 & 0 & V^\dagger & D & \cdots  \\
\vdots & \vdots & \vdots & \vdots & \ddots
\end{pmatrix},
\label{HNfull}
\end{eqnarray}
where we use $2 \times 2$ blocks
\begin{eqnarray*}
D &=&
\begin{pmatrix}
0 & \gamma_1 \kappa^\dagger \\
\gamma_1 \kappa & 0
\end{pmatrix} ,
\quad
V =
\begin{pmatrix}
0 & 0 \\
\gamma_1 & 0
\end{pmatrix} , \quad \kappa = \frac{\xi k_x + i k_y}{k_{\mathrm{c}}} \, .
\end{eqnarray*}
Here ${\bf k} = (k_x , k_y )$ is the wave vector measured from the center of valley $K_{\xi}$ with valley index $\xi = \pm 1$, and $k_{\mathrm{c}} = \gamma_1 / (\hbar v)$. Block $D$ describes intralayer nearest-neighbor hopping with velocity $v = (\sqrt{3}/2)a\gamma_0/\hbar$ and in-plane lattice constant $a$, block $V$ describes interlayer hopping with parameter $\gamma_1$ between successive $B_n$ and $A_{n+1}$ sites.
For numerical diagonalization of~(\ref{HNfull}) we use $\gamma_0 = 3.16\,$eV, $\gamma_1 = 0.381\,$eV~\cite{kuz09}, and $a = 2.46$\AA.

\subsection{Mean-field theory}

Electron-electron interactions are included within a mean-field Hartree-Fock approximation~\cite{min08,jung11,jung13} and, in particular, we follow the methodology applied to Bernal-stacked multilayer graphene in Refs.~\cite{yoon17,koshino17}. The total Hamiltonian is ${\hat H}_{\mathrm{tot}} = {\hat H} + {\hat V}_{\mathrm{MF}}$
where
\begin{eqnarray}
{\hat H} &=& \sum_{\mathbf{k} \sigma X \!X^{\prime}} H_{\mathbf{k} X\!X^{\prime}} c_{\mathbf{k} \sigma X}^\dagger c_{\mathbf{k} \sigma X^{\prime}}  ,  \label{H0} \\
{\hat V}_{\mathrm{MF}} &=& \sum_{\mathbf{k} \sigma X \!X^{\prime}} \left[ U_{X}^{(\mathrm{H})} \delta_{X\!X^{\prime}} + W_{\mathbf{k} \sigma X \!X^{\prime}} \right] c_{\mathbf{k} \sigma X}^\dagger c_{\mathbf{k} \sigma X^{\prime}} \label{VMF} .
\end{eqnarray}
Here $X = A_1,B_1,A_2,B_2, \ldots$ indexes the sublattices, $\sigma = 1,2,3,4$ is a flavor index combining spin $( \uparrow , \downarrow )$ and valley $(K_{+} , K_{-})$ degrees of freedom, 
and $c_{\mathbf{k} \sigma X}^\dagger$ and $c_{\mathbf{k} \sigma X}$ are creation and annihilation operators, respectively.
The non-interacting term ${\hat H}$ contains $H_{\mathbf{k} X\!X^{\prime}}$ which is a matrix element of~(\ref{HNfull}); the interaction term ${\hat V}_{\mathrm{MF}}$ consists of the Hartree $U^{(\mathrm{H})}$ and exchange $W$ potentials,
\begin{eqnarray}
U_{X}^{(\mathrm{H})} &=& \lim_{q \rightarrow 0} \sum_{X^{\prime}} 
V ({\bf q} ; z_{X} - z_{X^{\prime}}) n_{X^{\prime}} , \label{hartree} \\
\!\!\!\!\!\!\!\!\! W_{\mathbf{k} \sigma X \!X^{\prime}} &=& - \frac{1}{L^2} \sum_{{\bf k}^{\prime}} V({\bf k}-{\bf k}^{\prime} ; z_X - z_{X^{\prime}})
\langle c_{\mathbf{k}^{\prime} \sigma X^{\prime}}^\dagger c_{\mathbf{k}^{\prime} \sigma X} \rangle \! , \label{exchange}
\end{eqnarray}
where $n_X = (1/L^2) \sum_{\mathbf{k} \sigma} \langle c_{\mathbf{k} \sigma X}^{\dagger} c_{\mathbf{k} \sigma X} \rangle - n_0$, $L^2$ is the system area, $z_{X}$ is the vertical coordinate of sublattice $X$, and $V ({\bf q} ; z) = (2\pi e^2/[\epsilon_r q]) \exp (-q|z|)$
is the two-dimensional Fourier transform of the Coulomb potential~\cite{units}, $\epsilon_r$ is the dielectric constant and we use $d = 3.35$\AA \ for the interlayer separation.
Parameter $n_0$ represents the background density of positive charge and, in the charge neutral case considered here, it is determined by $\sum_{X} n_X = 0$.
In this case, the Hartree term~(\ref{hartree}) simplifies as
$U_{X}^{(\mathrm{H})} = - (2\pi e^2 / \epsilon_r) \sum_{X^{\prime}} |z_X - z_{X^{\prime}}|
n_{X^{\prime}}$.

The strength of the electronic interactions is characterized by the effective fine structure constant for graphene,
\begin{eqnarray}
\alpha_{\mathrm{g}} = \frac{e^2}{\epsilon_r \hbar v} .
\end{eqnarray}
For example, for $\epsilon_r = 2$ then $\alpha_{\mathrm{g}} \sim 1$.
However, the Hartree-Fock approximation tends to overestimate the strength of the exchange interaction by neglecting screening effects so we treat $\alpha_{\mathrm{g}}$ as a fitting parameter in the range $0 < \alpha_{\mathrm{g}} \leq 0.5$.

The  label $\sigma = 1,2,3,4$ takes four different values corresponding to flavors combining valleys and spin $(K_{+},\uparrow), (K_{+},\downarrow),(K_{-},\uparrow),(K_{-},\downarrow)$.
We neglect interactions between different valleys because they are described by the Coulomb interaction $V ({\bf q} ; z)$ with large wave vectors ${\bf q} \approx {\bf K}_{+} - {\bf K}_{-}$ and this approximation treats the four flavors on an equal footing. Within a given flavor, the exchange potential breaks chiral symmetry and charge density is transferred between layers. In addition, the four flavors have a certain relative configuration.
By beginning the iterative procedure with different initial exchange profiles, we find different self-consistent solutions, and we evaluate the total energy of each solution in order to determine the ground state.
For example, for the AF state, exchange for a given flavor has odd parity with respect to spatial inversion, and the four flavors are arranged in an antiferromagnetic configuration so that there is no net charge polarization and, hence, no cost in terms of Hartree energy. Thus this state has lower energy at charge neutrality than, say, a ferrimagnetic or ferromagnetic configuration of the four flavors.
Since our approximation treats the four flavors equally, it is unable to differentiate three distinct combinations~\cite{min08,zhang11,jung11} of spins and valleys within the antiferromagnetic configuration: layer-antiferromagnetic in which the polarization of flavors $(K_{+},\uparrow),(K_{-},\uparrow)$ is opposite to that of $(K_{+},\downarrow),(K_{-},\downarrow)$, quantum anomalous Hall when $(K_{+},\uparrow), (K_{+},\downarrow)$ are opposite to $(K_{-},\uparrow),(K_{-},\downarrow)$, or quantum spin Hall when $(K_{+},\uparrow)(K_{-},\downarrow)$ are opposite to $(K_{+},\downarrow),(K_{-},\uparrow)$.

The ground state at charge neutrality is found by numerically diagonalizing the Hamiltonian~(\ref{H0},\ref{VMF}) using an iterative procedure to determine a self-consistent solution taking the expectation values in $U^{(\mathrm{H})}$ and $W$ into account.
The summations over ${\bf k}$ are performed within a circle $k < k_{\star}$ around the $K$ point with a cutoff $k_{\star}$, and we choose $\hbar vk_{\star} \approx 1\,$eV~\cite{koshino17}. For fixed cutoff $k_{\star}$, the ground state is determined for different values of the system size $L$ ({\it i.e.} different densities of $k$ points), then the band gap is evaluated by extrapolation to $L \rightarrow\infty$. The numerical precision of our results is high so that uncertainties are negligible, and error bars are only shown in Section~\ref{disorder} when we study random disorder. Nevertheless, there are many sources of systematic uncertainty including the choice of cutoff $k_{\star}$, the omission of tight-binding parameters in the minimal model, and the value of the interaction parameter $\alpha_{\mathrm{g}}$. As described in Section~\ref{fbm}, we find very close agreement of band gap values at $\alpha_{\mathrm{g}} = 0.3$ with those obtained using DFT by Ref.~\cite{pamuk17} for $N=3$ to $N=8$ layers, and so we use $\alpha_{\mathrm{g}} = 0.3$ in subsequent sections.

Using the minimal model, the energy spectrum is isotropic around each $K$ point and the eigenstates of the non-interacting Hamiltonian~(\ref{HNfull}) at an arbitrary angle may be related to those at a specific angle by a stacking-dependent unitary transformation~\cite{min08,jang15}. We assume that the eigenstates of the interacting mean-field theory~(\ref{H0},\ref{VMF}) also satisfy this rotational transformation~\cite{yoon17}, allowing for the ${\bf k}$ summations to be performed in only one specific direction with the exchange interaction~(\ref{exchange}) being determined via an integration with respect to the polar angle of wave vector ${\bf k}^{\prime}$.
This simplification dramatically reduces the numerical cost of the calculations allowing for a study of multilayers with a large number of layers.

\subsection{Berry curvature and Chern number}

The non-interacting Hamiltonian~(\ref{HNfull}) obeys chiral symmetry~\cite{asboth16}: matrix elements only connect $A$ and $B$ sites (not $A$ to $A$ or $B$ to $B$) and chiral symmetry ensures particle-hole symmetry of the electronic spectrum. When the spectrum is gapped due to interactions, which generally break chiral symmetry, we use the wave functions in ${\bf k}$ space to determine the non-Abelian Berry curvature~\cite{fukui05} for the occupied valence bands.

In particular, at a discrete point in ${\bf k}$ space, we determine the wave functions $|u_n ({\bf k})\rangle$ for the valence bands with indices $n = 1 , .... , N$.
Following Ref.~\cite{fukui05}, we consider a lattice of cells with vertices at the discrete ${\bf k}$ points. For a cell centered at ${\bf k}$ and with $j$ vertices at ${\bf k}_1 , {\bf k}_2, \ldots , {\bf k}_{j-1} , {\bf k}_{j}$, the Berry flux $F_{{\bf k}}$~\cite{fukuicomment} is determined as
\begin{eqnarray}
F_{{\bf k}} = i \ln \left( U_{{\bf k}_1 , {\bf k}_2} U_{{\bf k}_2 , {\bf k}_3} \ldots U_{{\bf k}_{j-1} , {\bf k}_j} U_{{\bf k}_{j} , {\bf k}_1} \right) , \label{bflux}
\end{eqnarray}
where the link variables $U_{{\bf k}_a , {\bf k}_b}$ are evaluated for every side between vertices ${\bf k}_a$ and ${\bf k}_b$, and taken around the cell in the anti-clockwise direction.
For a side that connects vertices ${\bf k}_a$ and ${\bf k}_b$, the link variable $U_{{\bf k}_a , {\bf k}_b}$ is given~\cite{fukui05} by
\begin{eqnarray}
U_{{\bf k}_a , {\bf k}_b} &=& \frac{\det S_{{\bf k}_a , {\bf k}_b}}{| \det S_{{\bf k}_a , {\bf k}_b} |} , \\
( S_{{\bf k}_a , {\bf k}_b} )_{mn} &=& \langle  u_m ({\bf k}_a) | u_n ({\bf k}_b) \rangle .
\end{eqnarray}
Here, $S_{{\bf k}_a , {\bf k}_b}$ is a $N \times N$ matrix of scalar products between (valence) band wave functions with indices $m$, $n$.

Once the Berry flux~(\ref{bflux}) is determined, then the Berry curvature $\Omega_{\bf k}$ and Chern number $C$ are given by
\begin{eqnarray}
\Omega_{\bf k} = \frac{F_{{\bf k}}}{{\cal S}_{\bf k}} ; \qquad
C = - \frac{1}{2\pi} \sum_{\bf k} F_{{\bf k}} , \label{Csum}
\end{eqnarray}
where ${\cal S}_{\bf k}$ is the area of the cell at ${\bf k}$. In the continuous limit with many cells,
\begin{eqnarray}
C = - \frac{1}{2\pi} \int_{\mathrm{BZ}} \Omega ({\bf k}) d^2k . \label{Cint}
\end{eqnarray}
The sum in~(\ref{Csum}) and integral in~(\ref{Cint}) are taken over the first Brillouin zone (BZ).

Note that the cells do not need to have any particular shape (such as square), but ${\bf k}$ space should be covered by the cells and the cells should be sufficiently dense to achieve a convergent result for the Chern number~\cite{fukui05}. As we determine wave functions in the radial direction (from the valley center) and, then, relate them to wave functions at other angles by a rotation~\cite{min08,jang15}, we use cells that are parts of an annulus.
In our case, the Berry curvature is isotropic about the valley center, so the continuous case may be simplified as
\begin{eqnarray}
C_{\sigma} = - \int_0^{\infty} \Omega (k) k dk . \label{Cint2}
\end{eqnarray}
In the system we study, the Berry curvature is peaked at $k \alt k_c$ near a given valley and then falls to zero, and we integrate in the vicinity of a single valley (with a nominal upper limit at infinity) in order to determine the Chern number per spin-valley flavor $C_{\sigma}$~\cite{chernvalley}.

\begin{figure}[t]
\includegraphics[scale=0.2]{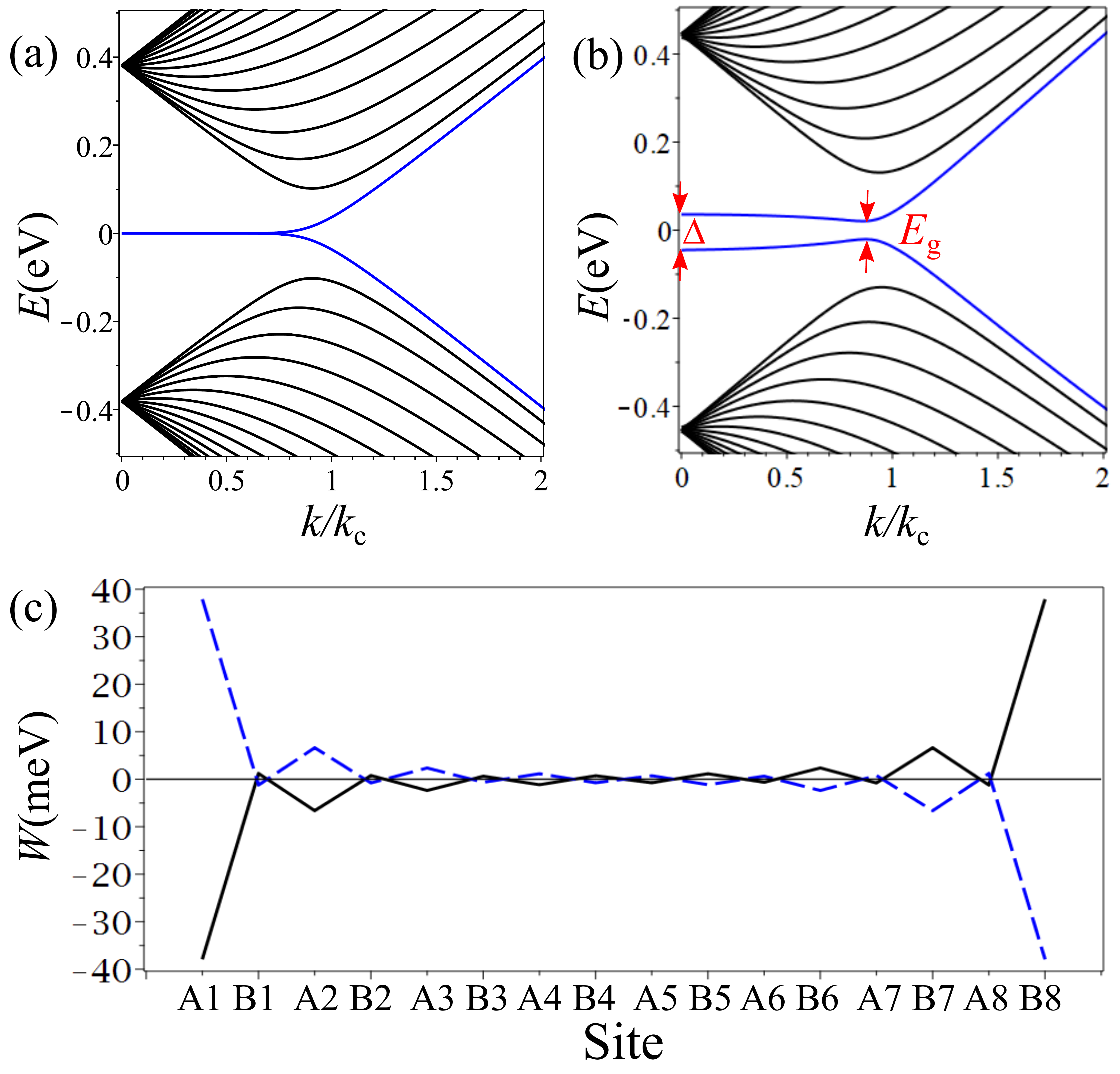}
\caption{Low-energy band structure of RMG with $N=16$ layers with (a) non-interacting electrons described by Hamiltonian~(\ref{HNfull}) and (b) interacting electrons described by the mean-field theory~(\ref{H0},\ref{VMF}) with interaction strength $\alpha_{\mathrm{g}} = 0.3$ and at zero temperature. Black lines are bulk bands, blue lines are surface bands for $k \alt k_{\mathrm{c}}$ where $\hbar vk_{\mathrm{c}} = \gamma_1$. $E_{\mathrm{g}}$ is the band gap (at $k \approx k_{\mathrm{c}}$), $\Delta$ is the order parameter (the separation of the surface bands at $k=0$).
(c) The values of the exchange potential $W_{k=0 \sigma X \!X}$ at the valley center $k=0$ and for each site $X^{\prime} = X$ for RMG with $N=8$ layers~\cite{exchangecomment}. The black solid line shows the exchange for two flavors, the blue dashed line is exchange for the other two flavors.
}\label{abc-bs1}
\end{figure}

\section{Rhombohedral graphene}\label{pristine}

The band structure of RMG with $N=16$ layers for a single spin-valley flavor is shown in Fig.~\ref{abc-bs1}(a,b) for non-interacting and interacting electrons, respectively, the latter as determined by numerical solving the mean-field theory~(\ref{H0},\ref{VMF}) for $2N$ bands, Eq.~(\ref{HNfull}).
The orbitals on the surface sites, $A_1$ and $B_N$, contribute to a pair of low-energy bands that are flat for $k \alt k_c$ where $k_c = \gamma_1 / \hbar v$. The wave vector $k_c$ corresponds to the point of the phase transition between non-trivial and trivial topological phases in the SSH model~\cite{su79,asboth16,heikkila11,xiao11}. For the interacting ground state, the band structure displays flavor degeneracy, Fig.~\ref{abc-bs1}(b).
The band gap, $E_{\mathrm{g}}$, is at finite wave vector $k$ between the surface state bands, this is the difference in energy between the conduction band minima and the valence band maxima. We also consider the separation of the surface state bands at $k = 0$ which we refer to as the order parameter $\Delta$. In principle, it is possible to have non-zero $\Delta$ even when $E_{\mathrm{g}} = 0$ and, generally,  $\Delta \geq E_{\mathrm{g}}$.

The exchange potential $W_{k=0 \sigma X \!X}$ at the valley center $k=0$ and for each site $X^{\prime} = X$ for RMG with $N=8$ layers is shown in Fig.~\ref{abc-bs1}(c)~\cite{exchangecomment}. For two flavors, the exchange is negative on $A$ sites and positive on all $B$ sites, with large magnitude on the surface sites $A_1$ and $B_N$ (solid black line); for the other two flavors, the exchange has an inverted profile (dashed blue line).
We refer to this as being the odd antiferromagnetic state because the exchange potential has odd parity within each flavor, and the four flavors are arranged in an antiferromagnetic configuration.

Before discussing the layer number $N$ and temperature $T$ dependence of $E_{\mathrm{g}}$ and $\Delta$ arising from the numerical calculations, we consider a very simple two-band model that may be solved analytically in order to develop a qualitative picture of the AF state.

\subsection{Two band model}

We consider a mean-field Hamiltonian~\cite{mccann06,novo06} for each of the four $\sigma$ flavours with two sublattices $A_1$ and $B_N$, and order parameter $\Delta = 2|w|$ due to the exchange interaction $w$ which breaks sublattice symmetry:
\begin{eqnarray}
H_2 = \left(
      \begin{array}{cc}
        E_0 + w &   -\gamma_1 (- \kappa^{\dagger})^N  \\
           -\gamma_1 (- \kappa)^N & E_0 - w \\
      \end{array}
    \right) , \label{twoband}
\end{eqnarray}
where $N \geq 2$.
Note that similar two-band models have been considered in similar contexts previously~\cite{min08,koshino10}.
The eigenvalues and eigenstates of $H_2$ may be written as
\begin{eqnarray}
\!\!\!\!\!\! E_{s} &=& E_0 + s\sqrt{w^2 + \gamma_1^2 (k/k_{\mathrm{c}})^{2N}} , \label{tmev} \\
\psi_{s} &=& \! \frac{1}{\sqrt{2(E_s-E_0)(E_s-E_0+w)}} \!\!  \left( \!\!\!
                                              \begin{array}{c}
                                                E_s - E_0 + w \\
                                                - \gamma_1 (-\kappa )^N   \\
                                              \end{array}
                                            \!\!\! \right) \!\! , \label{tmes}
\end{eqnarray}
where $s = \pm 1$ indexes conduction and valence bands, and $k = |{\bf k}| \equiv \sqrt{k_x^2+k_y^2}$.
We will show this is a self-consistent solution under the approximation that
we only take into account the contribution exactly at the valley center ($k = 0$)
in the exchange~(\ref{exchange}).
This means that parameter $w$ is independent of $k$, and that the off-diagonal in sublattice exchange potential $W_{\mathbf{k} \sigma A1 BN}$ is zero because the summation over all ${\bf k}^{\prime}$ includes a factor such as $\exp (iN\phi^{\prime})$ arising from the chiral wave functions~(\ref{tmes}), where $\phi^{\prime}$ is the polar angle of the wave vector ${\bf k}^{\prime}$.

Considering the diagonal in sublattice exchange potential $W_{\mathbf{k} \sigma A1 A1}$~(\ref{exchange}), then
\begin{eqnarray}
w =  - \frac{2\pi \hbar v \alpha_{\mathrm{g}}}{L^2} \sum_{s = \pm 1} \sum_{{\bf k}^{\prime}}
\frac{f(E_s)}{|{\bf k}^{\prime}|}
 \frac{w}{2(E_{s}-E_0)} , 
\end{eqnarray}
where $f(E_s) = 1/(\exp [(E_s - E_0)/(k_BT)] + 1)$ is the Fermi-Dirac distribution, $E_0$ is the chemical potential, $k_B$ is Boltzmann's constant and $T$ is absolute temperature.
Then, the equation for the order parameter $\Delta = 2|w|$ is
\begin{eqnarray}
\Delta &=& 2\gamma_1 \left[ \frac{\alpha_{\mathrm{g}}}{2} g_N \!\! \left(\frac{\Delta}{k_B T}\right) \right]^{N/(N-1)} , \label{finitet}\\
g_N (x) &=& \int_{0}^{\infty} \!\! \frac{dy}{\sqrt{1+y^{2N}}} \frac{\sinh \left( \tfrac{x}{2} \sqrt{1+y^{2N}} \right)}{1 + \cosh \left( \tfrac{x}{2} \sqrt{1+y^{2N}} \right)} .
\end{eqnarray}
At zero temperature, $g_N (\infty)$ is simply a number, and the order parameter is explicitly given by
\begin{eqnarray}
\Delta (T=0) &=& 2 \gamma_1 \left( \frac{\alpha_{\mathrm{g}} g_N (\infty) }{2} \right)^{N/(N-1)} , \\
g_N (\infty) &=& \frac{1}{\sqrt{\pi}} \Gamma \!\left( 1 + \frac{1}{2N} \right) \Gamma \!\left( \frac{1}{2} - \frac{1}{2N} \right) ,
\end{eqnarray}
where $\Gamma (x)$ is the gamma function.
For bilayer graphene, $N = 2$, then $g_2 (\infty) = 1.854$ and $\Delta (T=0) = 1.719 \alpha_{\mathrm{g}}^2 \gamma_1$, whereas, for $N \gg 1$, then $g_N (\infty) = 1$ and $\Delta (T=0) = \alpha_{\mathrm{g}} \gamma_1$.
For finite $T$, the temperature dependence of the order parameter~(\ref{finitet}) is similar to the self-consistent equation for the magnetization in the Weiss mean-field approximation~\cite{kittel} and, for $N \gg 1$, we find that the critical temperature is given by $k_{\mathrm{B}}T_{\mathrm{c}} = \alpha_{\mathrm{g}} \gamma_1/4$.

\begin{figure}[t]
\includegraphics[scale=0.21]{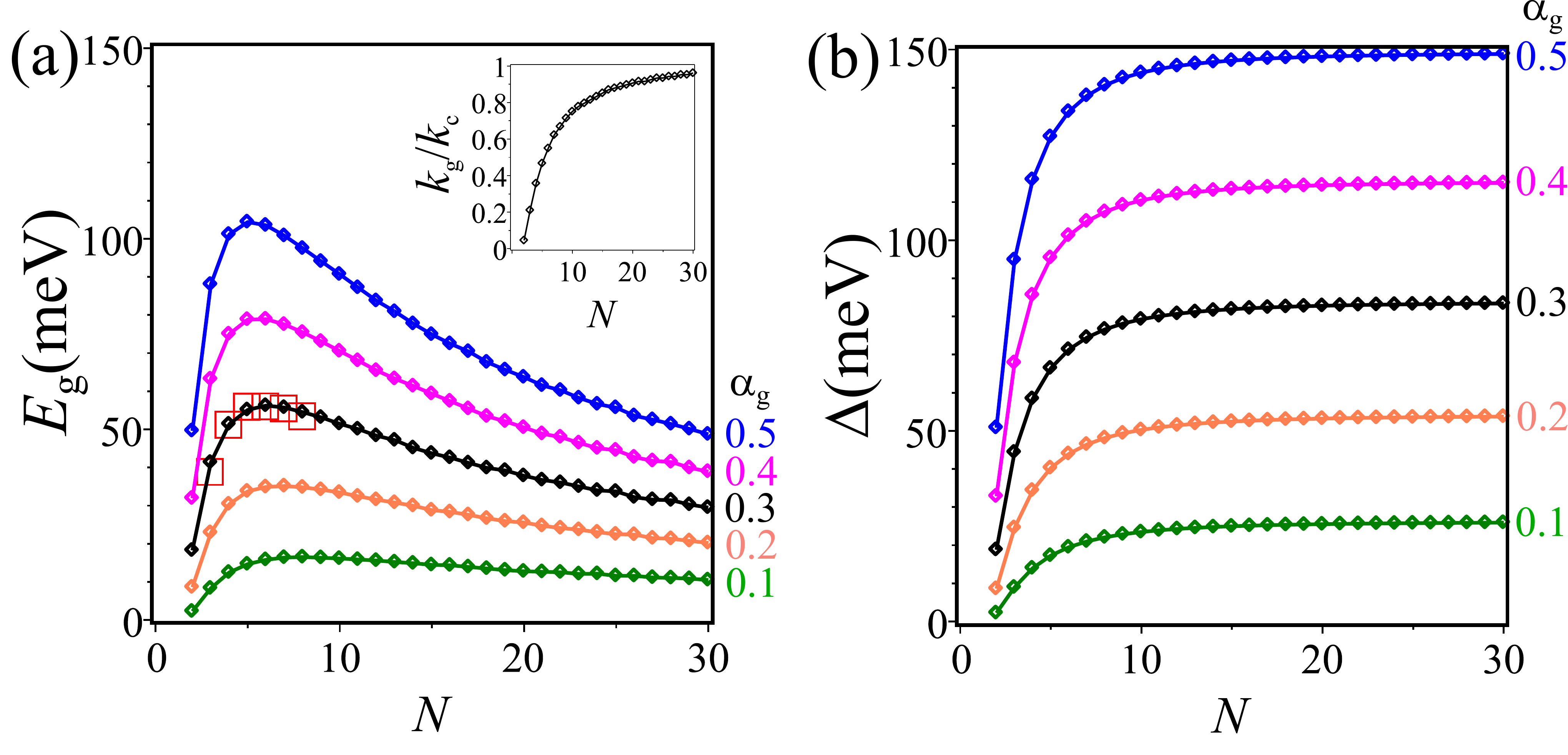}
\caption{The band gap of rhombohedral multilayer graphene at zero temperature as a function of layer number $N$ for different values of interaction strength $\alpha_{\mathrm{g}}$ showing (a) the band gap $E_{\mathrm{g}}$ and (b) the order parameter $\Delta$. Solid lines are a guide for the eye. Red squares in (a) show the results of Ref.~\cite{pamuk17} for $N=3$ to $N=8$ obtained using density functional theory. The inset of (a) shows the position of the band gap $k_{\mathrm{g}}$ in units of $k_{\mathrm{c}}$ as a function of $N$ for $\alpha_{\mathrm{g}} = 0.3$.
}\label{abc-gap1}
\end{figure}

\subsection{Full band model}\label{fbm}

We now discuss the results of the numerical calculation to solve the mean-field theory~(\ref{H0},\ref{VMF}) for $2N$ bands. 
The band gap, $E_{\mathrm{g}}$, and the order parameter (the band separation at $k=0$), $\Delta$, are plotted in Fig.~\ref{abc-gap1} as a function of layer number $N$ for different interaction strengths $\alpha_{\mathrm{g}}$.
Red squares in Fig.~\ref{abc-gap1}(a) show the results of Ref.~\cite{pamuk17} for $E_{\mathrm{g}}$ for $N=3$ to $N=8$ obtained using DFT. Our results are in qualitative agreement and, by choosing $\alpha_{\mathrm{g}} = 0.3$, close quantitative agreement with those of Ref.~\cite{pamuk17}: the band gap, $E_{\mathrm{g}}$, grows for small $N$, until it peaks around $N=6$ and, then, falls for larger $N$. The order parameter, $\Delta$, also grows for small $N$, but it saturates for larger $N$ values (as in the simple two-band model).

The increase of $E_{\mathrm{g}}$ and $\Delta$ for small $N$ is attributed to an increasing density of states of progressively flatter and flatter bands~\cite{myhro18,pamuk17}.
The decrease of $E_{\mathrm{g}}$ at large $N$ is largely due to the non-interacting band structure, as described by Hamiltonian~(\ref{HNfull}), in that the position of the band gap moves from $k \approx 0$ for $N=2$ to $k \approx k_{\mathrm{c}}$ for $N \gg 1$ as shown in the inset of Fig.~\ref{abc-gap1}(a); the bulk gap [i.e. between the bulk bands shown in black in Fig~\ref{abc-bs1}(b)] closes at $k \approx k_{\mathrm{c}}$ for $N \gg 1$~\cite{henni16,slizovskiy19}.

\begin{figure}[t]
\includegraphics[scale=0.2]{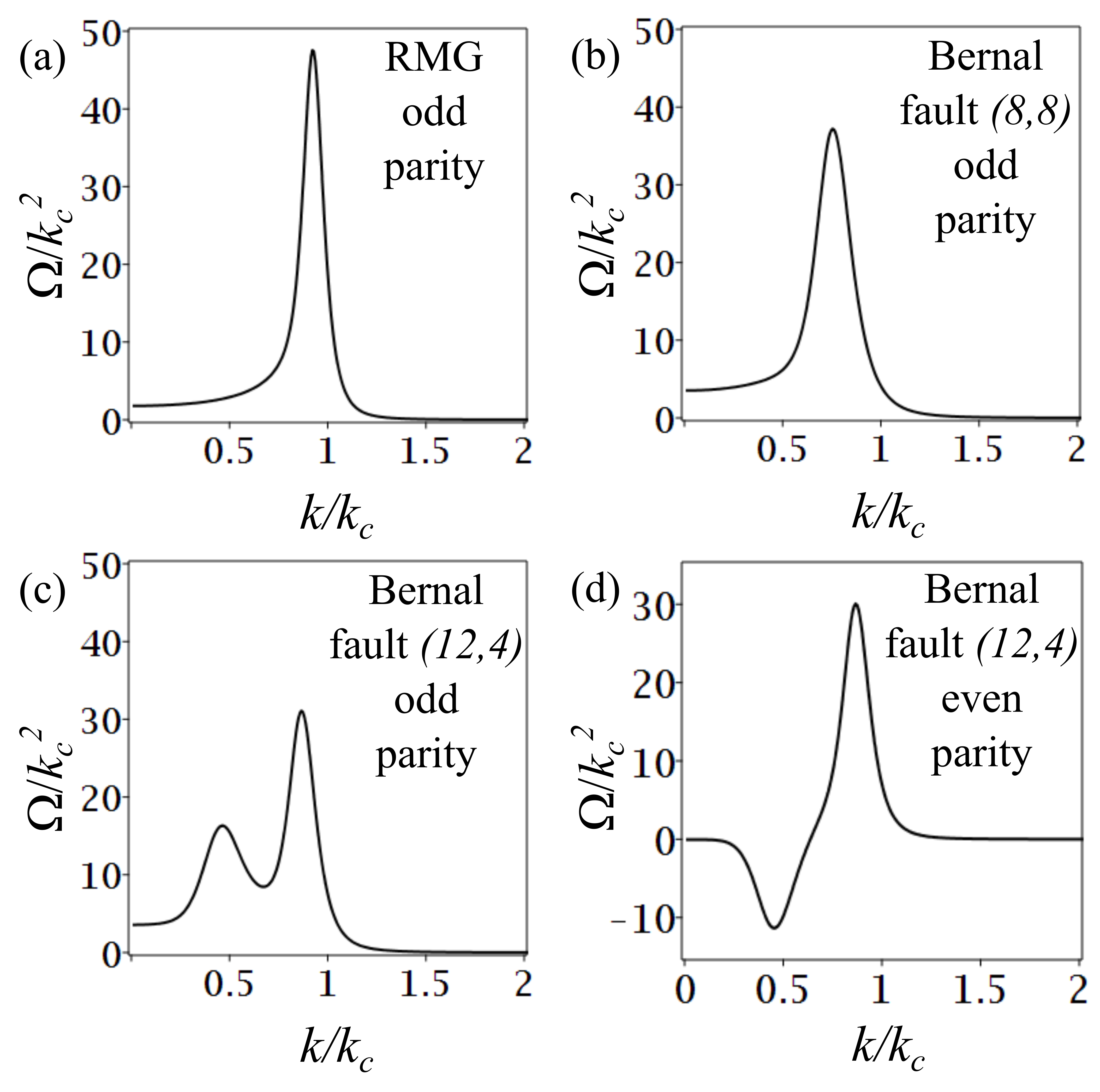}
\caption{Berry curvature $\Omega (k)$ as a function of the magnitude of the wave vector $k$ plotted from the valley center (the Berry curvature is isotropic in the minimal model) for $N=16$, $\alpha_{\mathrm{g}} = 0.3$ and $T=0\,$K. (a) is for pristine RMG in the odd parity ground state, (b) is for RMG with a Bernal stacking fault at its center in the odd parity ground state, (c) is for an off-centre Bernal stacking fault in the odd parity ground state, (d) is for an off-centre Bernal stacking fault in an even parity  state (which is not the ground state).
}\label{berry1}
\end{figure}

For the odd parity ground state, the Berry curvature $\Omega (k)$ is plotted in Fig.~\ref{berry1}(a) as a function of the magnitude of the wave vector $k$ plotted from the valley center (the Berry curvature is isotropic in the minimal model) for $N=16$. The position of the maximum in $\Omega (k)$ is given by $k_{\mathrm{max}} \approx s_N k_c (\Delta /\gamma_1)^{1/N}$, $s_N = ((N-1)/(2N+4))^{1/2N}$, which moves from $k=0$ to $k = k_c$ as $N$ increases~\cite{xiao07,xiao10,zhang11,slizovskiy19}. The integral of $\Omega (k)$ with respect to the wave vector area also increases with $N$, as characterized by the Chern number per flavor which has magnitude $N/2$~\cite{fukui05,asboth16,chernvalley}.

\begin{figure}[t]
\includegraphics[scale=0.21]{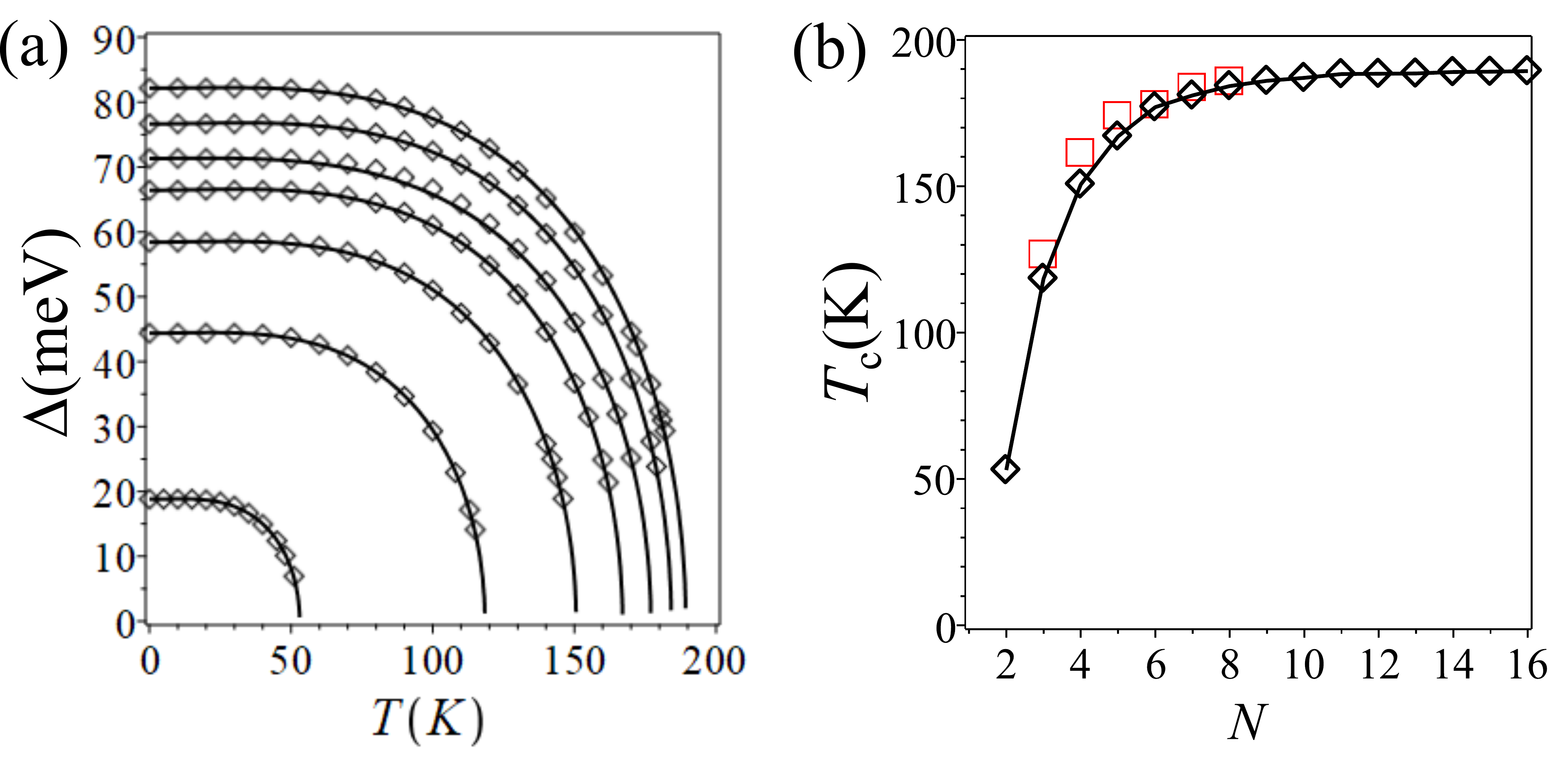}
\caption{(a) the order parameter $\Delta (T)$ of rhombohedral multilayer graphene as a function of temperature $T$ for interaction strength $\alpha_{\mathrm{g}} = 0.3$ and layer number $N = 2,3,4,5,6,8,16$ from bottom to top. Solid lines are fits according to Eq.~(\ref{tcfit}). (b) Critical temperature $T_c$ as a function of layer number $N$ for interaction strength $\alpha_{\mathrm{g}} = 0.3$ obtained using the fit~(\ref{tcfit}). The solid line is a guide to the eye. Red squares show the results of Ref.~\cite{pamuk17} for $N=3$ to $N=8$ obtained using density functional theory.
}\label{abc-tc1}
\end{figure}

Finite temperature is taken into account through the Fermi-Dirac distribution in the mean field theory Eqs.~(\ref{hartree},\ref{exchange}) and $\Delta (T)$ is shown in Fig.~\ref{abc-tc1}(a) for different $N$ values and interaction strength $\alpha_{\mathrm{g}} = 0.3$. We fit $\Delta (T)$ to a form suggested in Ref.~\cite{pamuk17},
\begin{eqnarray}
\frac{\Delta (T)}{\Delta (0)} &=& \bigg[ A \left(1 - \frac{T}{T_{\mathrm{c}}} \right) + (3-2A)\left(1 - \frac{T}{T_{\mathrm{c}}} \right)^2 \nonumber \\
&& \qquad + (A-2)\left(1 - \frac{T}{T_{\mathrm{c}}} \right)^3 \bigg]^{1/2} , \label{tcfit}
\end{eqnarray}
with $A$ and $T_{\mathrm{c}}$ as temperature independent fitting parameters. As shown in Fig.~\ref{abc-tc1}(a), the quality of this fit is generally excellent. The resulting layer dependence of the critical temperature $T_{\mathrm{c}} (N)$ is shown in Fig.~\ref{abc-tc1}(b). Red squares in Fig.~\ref{abc-tc1}(b) show the results of Ref.~\cite{pamuk17} for $N=3$ to $N=8$ obtained using density functional theory (DFT); our results are in close agreement. As in the simple two-band model, $T_{\mathrm{c}}$ saturates at a finite value for $N \gg 1$, and, for $\alpha_{\mathrm{g}} = 0.3$, the numerical calculation gives $\Delta (T=0) / (k_{\mathrm{B}}T_{\mathrm{c}}) \sim 5$ which is slightly larger than $\Delta (T=0) / (k_{\mathrm{B}}T_{\mathrm{c}}) = 4$ for the two-band model.
Note that we fit using $\Delta (T)$ rather than $E_g (T)$ (as in Ref.~\cite{pamuk17}) because $E_g$ values are determined at finite $k$, and the use of a finite number of discrete $k$ points introduces slightly more uncertainty (than the determination of $\Delta (T)$ which is always at $k=0$).

\section{Interlayer disorder}\label{disorder}

In this section, we consider the influence of interlayer disorder on the interacting mean-field state in RMG, {\it i.e.} we preserve translational invariance within each graphene layer, but take into account random tight-binding parameters in the perpendicular-to-layer direction. This has a close analogy with studies of the SSH model~\cite{su79,asboth16} wherein the effects on the zero-energy edge states of chiral-symmetry-preserving or breaking disorder are considered~\cite{mondragonshem14,liu18,perezgonzalez19,jurss19,scollon20}. A major difference here is that we consider the influence of disorder on the interacting mean-field state in which the exchange potential has already broken chiral symmetry and gapped the spectrum.

\begin{figure}[t]
\includegraphics[scale=0.21]{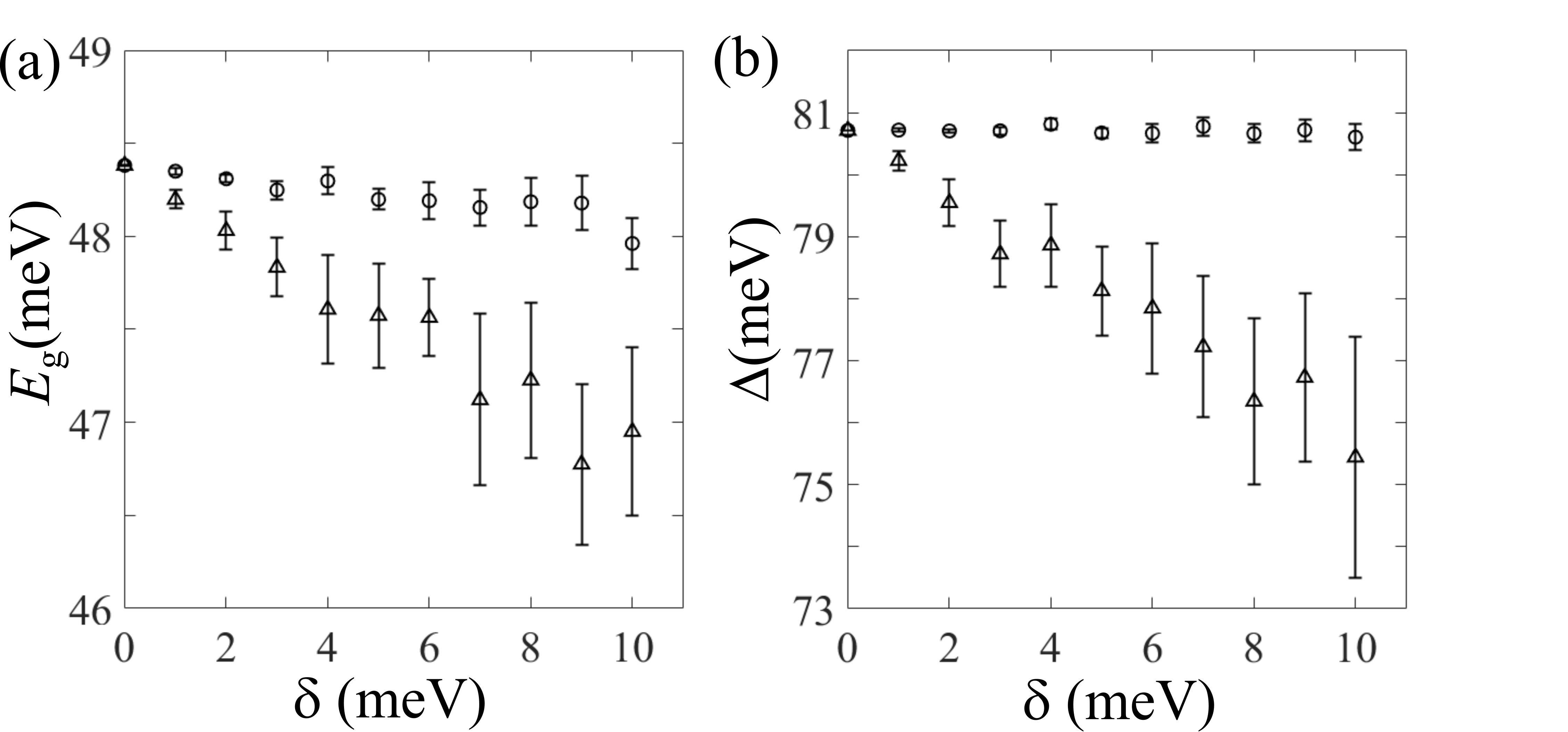}
\caption{The mean band gap of rhombohedral multilayer graphene with $N=12$ layers in the odd parity antiferromagnetic state at zero temperature and $\alpha_{\mathrm{g}} = 0.3$ as a function of disorder strength $\delta$ showing (a) the mean band gap $E_{\mathrm{g}}$ and (b) the mean order parameter $\Delta$. Triangles show data for random layer energies, circles show data for random interlayer coupling. Mean values and error bars are found by averaging over twenty disorder realizations. Note that the scale on the vertical axes is offset from zero, and is different in (a) and (b).
}\label{disorderplot}
\end{figure}

We take into account two types of disorder. The first is due to random layer energies which break chiral symmetry. For a given realization of disorder, the diagonal elements of the non-interacting Hamiltonian~(\ref{HNfull}) take values $H_{A_nA_n} = H_{B_nB_n} = \delta_n$ for layer index $n = 1,2,3,\ldots , N$ where each $\delta_n$ takes a random value uniformly distributed in the range $[-\delta , \delta]$ for disorder strength $\delta$. We consider weak disorder up to $\delta = 10\,$meV so that $\delta \ll \{ E_g , \gamma_1 \}$ for typical values of the band gap $E_g$. Figure~\ref{disorderplot}(a) shows the mean band gap $E_g$ for $N=12$ layers as a function of disorder strength $\delta$ for the odd parity antiferromagnetic state, Fig.~\ref{disorderplot}(b) shows the mean order parameter. Each data point (triangles) is an average over twenty different realizations of disorder; error bars increase with disorder in Fig.~\ref{disorderplot} because the standard deviation increases while the number of realizations is constant. The mean values of both $E_g$ and $\Delta$ decrease with disorder $\delta$, although they appear to be quite robust for weak disorder. We restrict the study to weak disorder values because the ground state will change ({\it e.g.} to an odd ferrimagnetic state as modeled in bilayer graphene~\cite{jung11}) for certain realizations at higher disorder.

The second type of disorder is due to random values of the interlayer coupling $\gamma_1$ which preserve chiral symmetry. For a given realization of disorder, elements of the non-interacting Hamiltonian~(\ref{HNfull}) describing interlayer coupling take random values, {\it i.e.} $H_{B_1A_2} = H_{A_2B_1} = \gamma_1 + \delta_1$, $H_{B_2A_3} = H_{A_3B_2} = \gamma_1 + \delta_2$, {\it etc.}, $n = 1,2,3,\ldots , N-1$, where each $\delta_n$ takes a random value uniformly distributed in the range $[-\delta , \delta]$ for disorder strength $\delta$. Figure~\ref{disorderplot} shows the dependence of the mean values of $E_g$ and $\Delta$ for $N=12$ layers as a function of disorder strength $\delta$ for the odd parity antiferromagnetic state, each data point (circles) is an average over twenty different realizations of disorder. We find that the mean value of $\Delta$ is not affected by disorder (within the error bars), and that disorder slightly reduces the mean value of $E_g$ for the weak disorder values we consider. This is in line with studies of the SSH model~\cite{mondragonshem14,liu18,perezgonzalez19,jurss19,scollon20} where one expects chiral-preserving disorder to have a negligible effect on the zero-energy edge states, although the exchange interaction has already broken chiral symmetry in the interacting mean-field state considered here.

The influence of disorder may be understood by considering the form of the two-band Hamiltonian~(\ref{twoband}). For random layer energies, the energies of the outer layers would appear directly in the two-band Hamiltonian as random diagonal elements for $H_{A_1A_1}$ and $H_{B_NB_N}$, having a direct impact on the exchange potential and the band gap in the form of random numbers $\delta_1$ and $\delta_N$. Interlayer coupling, however, appears in the off-diagonal term as $H_{A_1B_N} = - \gamma_1 (-\kappa^{\dagger})^N = - ( - \hbar v [\xi k_x - i k_y] )^N \gamma_1^{1-N}$, {\it i.e.} the connection between the surface states involves a product of the $N-1$ parameters $\gamma_1 + \delta_n$ for $n = 1,2,3,\ldots , N-1$. For $N \gg 1$, the system self-averages so that the effect of random values $\delta_n$ is negligible for the low-energy bands.

\begin{figure}[t]
\includegraphics[scale=0.2]{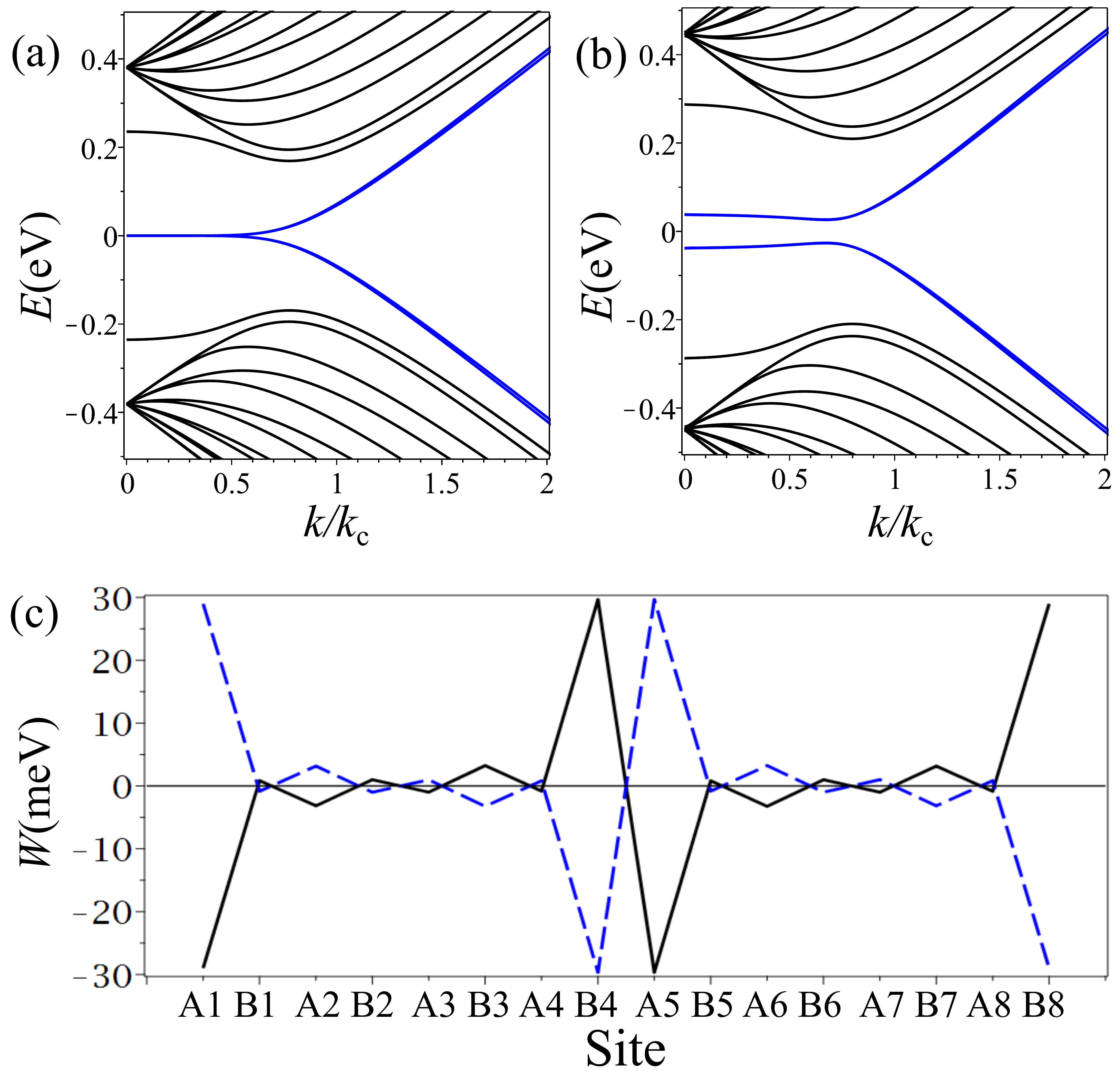}
\caption{Low-energy band structure of RMG with $N=16$ layers and a Bernal stacking fault at its center with (a) non-interacting electrons described by Hamiltonian~(\ref{HNfull}) and (b) interacting electrons described by the mean-field theory~(\ref{H0},\ref{VMF}) with interaction strength $\alpha_{\mathrm{g}} = 0.3$ and zero temperature. Black lines are bulk bands, blue lines are almost doubly-degenerate low-energy bands for $k \alt k_{\mathrm{c}}$ where $\hbar vk_{\mathrm{c}} = \gamma_1$ ($8$-fold degenerate taking into account spin and valley degrees of freedom). 
(c) The values of the exchange potential $W_{k=0 \sigma X \!X}$ at the valley center $k=0$ and for each site $X^{\prime} = X$ for RMG with $N=8$ layers and a Bernal stacking fault at its center~\cite{exchangecomment}. The black solid line shows the exchange for two flavors, the blue dashed line is exchange for the other two flavors.
}\label{abfault-bs1}
\end{figure}

\section{Bernal stacking fault in rhombohedral graphene}\label{bernal}

Stacking faults have been considered previously in graphene multilayer systems~\cite{arovas08,koshino13} and in RMG in particular~\cite{taut14,garciaruiz19,shi20}. Single stacking faults, {\it e.g.} a Bernal fault or a twin boundary fault, Fig.~\ref{lattice}, within a large RMG system are interesting because they introduce additional flat bands into the energy spectrum. Here we focus on the properties of the interacting ground state at half filling.

\subsection{Band structure of non-interacting electrons}

In order to describe the influence of a stacking fault on the low-energy band structure, we begin by considering non-interacting electrons. The number of zero energy states at $k=0$ is determined by the stacking structure of the multilayer, particularly the degree of hybridization of $p_z$ orbitals (one per site) caused by interlayer coupling $\gamma_1$ with a neighboring site in an adjacent layer directly above or below. For $n$ atoms coupled in a vertical line by interlayer coupling $\gamma_1$, even $n$ ({\it e.g.} a dimer) contributes $n$ bulk bands, but no zero energy states at $k=0$. For odd $n$ ({\it e.g.} a monomer not directly connected to a neighbor in an adjacent layer or a trimer), there are $n-1$ bulk bands and one zero energy state at $k=0$~\cite{min08b}.

For RMG, all sites are part of a dimer apart from $A_1$ and $B_N$ at the surfaces which are monomers: hence there are two zero energy states as shown in Fig.~\ref{abc-bs1}. For a Bernal-stacked multilayer with $N$ layers, there is one $N$-mer contributing one zero energy state if $N$ is odd and in addition there are $N$ monomers. Thus, overall, there are $N$ zero energy states if $N$ is even, $N+1$ if $N$ is odd.

We consider RMG with a single Bernal stacking fault, as illustrated in Fig.~\ref{lattice} (central panel) for $N = 7$ layers. Specifically, we use integers $(m,n)$ to denote
a rhombohedral section with $m$ layers and sites $A_1$, $B_1$, \ldots , $A_m$, $B_m$ connected by a Bernal stacking fault to a rhombohedral section with $n$ layers and sites $A_{m+1}$, $B_{m+1}$, \ldots $A_N$, $B_N$ where the total layer number is $N = m+n$ and $m \geq 2$, $n \geq 2$.
Thus, the example in the central panel of Fig.~\ref{lattice} is a $(3,4)$ Bernal stacking fault.
At the stacking fault, there are four vertically connected atomic sites (sites $B_2$, $A_3$, $B_4$, $A_5$ in Fig.~\ref{lattice}) which make up a $4$-mer; this is even, so it contributes $4$ bulk bands, but no zero energy states at $k=0$. Rather, the zero energy states arise from the sites not directly connected to a neighbor in an adjacent layer, namely $A_1$, $B_m$, $A_{m+1}$, $B_N$ (sites $A_1$, $B_3$, $A_4$, $B_7$ in Fig.~\ref{lattice}), so there are four low-energy states per spin and valley flavor.

The low-energy bands of non-interacting electrons for $N=16$ layers with a Bernal stacking fault at the center $n=m=8$ are shown in Fig.~\ref{abfault-bs1}(a).
For $m,n \gg 1$, the non-interacting low-energy bands behave almost as if they arise from two disconnected pieces of RMG with $m$ and $n$ layers, respectively. This may be understood by deriving an effective low-energy four band Hamiltonian, following the procedure described previously for bilayer graphene~\cite{mccann06,mccann13}, in a basis of orbitals on $A_1$, $B_m$, $A_{m+1}$, $B_N$ sites. For $k \ll k_c$ and $E \ll \gamma_1$, we find
\begin{widetext}
\begin{eqnarray}
H_{\mathrm{Bernal}}^{(m,n)} = \gamma_1\left(
           \begin{array}{cccc}
             0 & - (-\kappa^{\dagger})^m & 0 & (-\kappa^{\dagger})^{m+n-2} \\
             - (-\kappa)^m & 0 & - c_{mn} (k/k_{\mathrm{c}})^{2(\ell-1)}\kappa^2 & 0 \\
             0 & - c_{mn} (k/k_{\mathrm{c}})^{2(\ell-1)}(\kappa^{\dagger})^2 & 0 & - (-\kappa^{\dagger})^n \\
             (-\kappa)^{m+n-2} & 0 & - (-\kappa)^n & 0 \\
           \end{array}
         \right) , \label{hbernal}
\end{eqnarray}
\end{widetext}
where $c_{mn} = (1+\delta_{mn})/2$ and $\ell = \mathrm{min} (m,n)$. As the Hamiltonian is chiral, every matrix element between two A sites or between two B sites is zero. For the non-zero elements (between A and B sites), we keep only the leading terms in $k/k_c$.

The diagonal $2 \times 2$ blocks in Eq.~(\ref{hbernal}) describe isolated RMG sections with $m$ and $n$ layers; the off-diagonal $2 \times 2$ blocks describe coupling between them. In particular, term $(-\kappa)^{m+n-2}$ describes effective coupling between the $A_1$ and $B_N$ sites which are on opposite surfaces of the sample and this is very small for $N \gg 1$ and $k/k_c \ll 1$. Term $- c_{mn} (k/k_{\mathrm{c}})^{2(\ell-1)}(\kappa^{\dagger})^2$ describes effective coupling between the $B_m$ and $A_{m+1}$ sites. Although they are on adjacent layers, this coupling is of order $(k/k_c)^{2\ell}$ which is also very small for $\ell \gg 1$ and $k/k_c \ll 1$. Weak effective coupling between the $B_m$ and $A_{m+1}$ sites arises from the fact that the Bernal stacking fault consists of four vertically coupled sites ($B_2$, $A_3$, $B_4$, $A_5$ in Fig.~\ref{lattice}). Their effective coupling in the basis of Eq.~(\ref{hbernal}) is described by inverting the $4 \times 4$ matrix of hopping within a $4$-mer,
\begin{eqnarray*}
\left(
           \begin{array}{cccc}
             0 & 1 & 0 & 0 \\
            1 & 0 & 1 & 0 \\
             0 & 1 & 0 & 1 \\
             0 & 0 & 1 & 0 \\
           \end{array}
         \right)^{-1}
=
\left(
           \begin{array}{cccc}
             0 & 1 & 0 & -1 \\
            1 & 0 & 0 & 0 \\
             0 & 0 & 0 & 1 \\
             -1 & 0 & 1 & 0 \\
           \end{array}
         \right) ,
\end{eqnarray*}
which has an exactly zero matrix element between the second and third components. The reason that matrix element $- c_{mn} (k/k_{\mathrm{c}})^{2(\ell-1)}(\kappa^{\dagger})^2$ in Eq.~(\ref{hbernal}) is not also identically zero is that this small contribution arises from a slight rotation of the low-energy basis states that is required to preserve their normalization~\cite{mccann13}.

Since the four band Hamiltonian~(\ref{hbernal}) is chiral, the energy spectrum of non-interacting electrons displays electron-hole symmetry and the band energies $E$ may be determined as the solution of a quadratic equation:
\begin{eqnarray*}
( E /\gamma_1 )^2 &=& \frac{1}{2} \beta_{\mathrm{B}} (k/k_c) \pm \frac{1}{2} \sqrt{ \beta_{\mathrm{B}}^2 (k/k_c) - 4 \eta_{\mathrm{B}} (k/k_c)} , \\
\beta_{\mathrm{B}} (x) &=& x^{2m} + x^{2n} + x^{2m+2n-4} + c_{mn}^2 x^{4\ell} , \\
\eta_{\mathrm{B}} (x) &=& x^{2m+2n} + 2 c_{mn} x^{2m+2n+2\ell -2} + c_{mn}^2 x^{2m+2n+4\ell - 4} .
\end{eqnarray*}
Of particular interest is when the fault lies exactly in the center of a long RMG system: $n=m = N/2$, $c_{mn} = 1$, and $\ell = m = N/2$ with $N \gg 1$. Then, the stacking fault (as described by the off-diagonal $2 \times 2$ blocks in Eq.~(\ref{hbernal})) connects the two RMG sections and breaks the degeneracy of their spectra: $E \approx \pm \gamma_1 \left[ (k/k_c)^{N/2} \pm \tfrac{1}{2} (k/k_c)^{N-2} \right]$ for $k \ll k_c$.

\begin{figure}[t]
\includegraphics[scale=0.21]{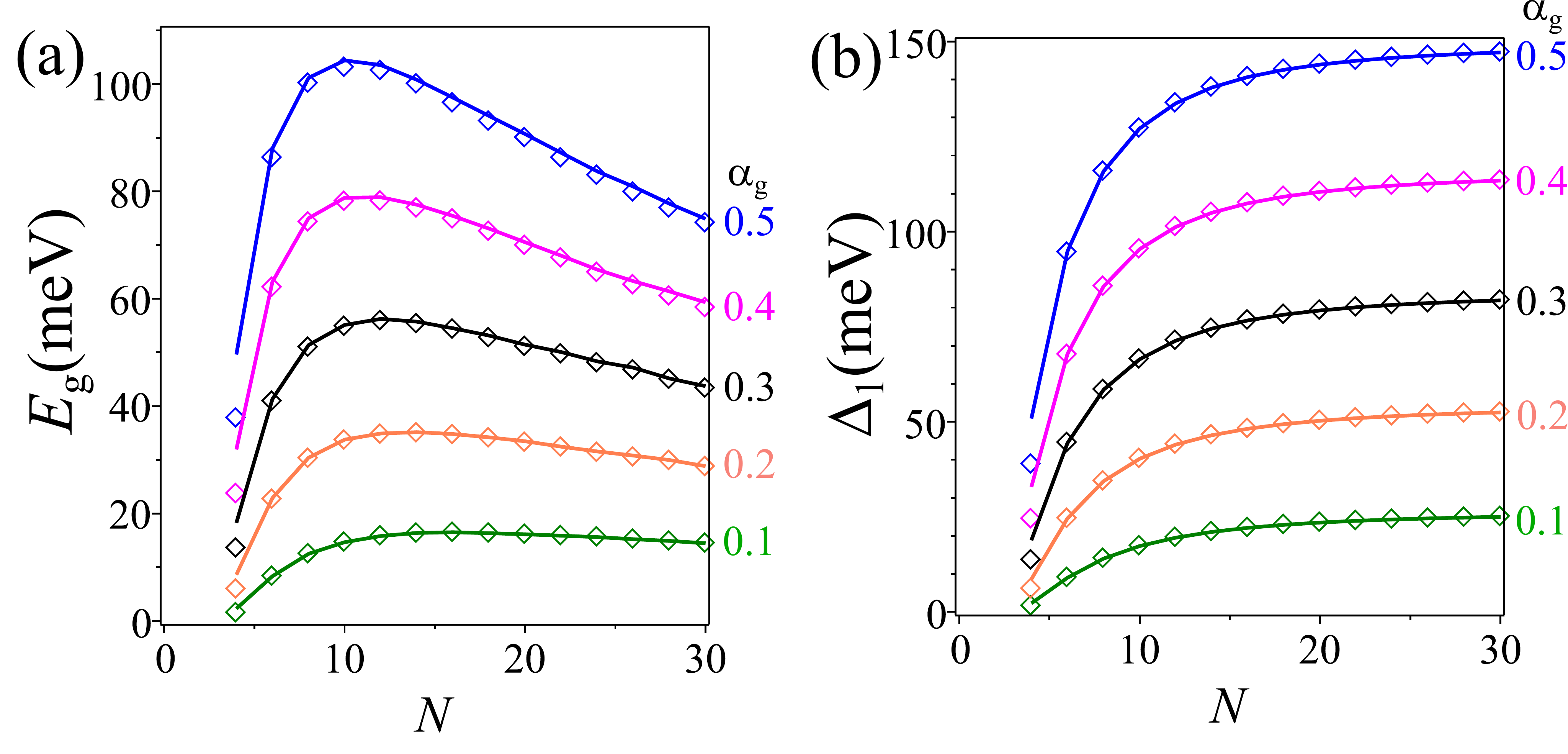}
\caption{The band gap of RMG with a Bernal stacking fault at its center, for even $N$, at zero temperature as a function of layer number $N$ for different values of interaction strength $\alpha_{\mathrm{g}}$ showing (a)  the band gap $E_{\mathrm{g}}$ and (b) the order parameter $\Delta_1$. Points are data for the system with the stacking fault, solid lines are data from Fig.~\ref{abc-gap1} for a single RMG section with $m = N/2$ layers.
}\label{abfault-gap1}
\end{figure}

\subsection{Numerical mean-field theory}

The numerical mean-field theory calculations proceed as for pristine RMG, except that the stacking fault is taken into account by a different position of the interlayer coupling $\gamma_1$ in the non-interacting Hamiltonian~(\ref{HNfull}).
At zero temperature, we find a number of self-consistent solutions with different exchange profiles including both even and odd parity within each flavor, and ferromagnetic, ferrimagnetic or antiferromagnetic arrangements of flavors. However, on determining the total energy of each, we find that the ground state is the odd antiferromagnetic state, a generalization of the ground state in pristine RMG (discussed in Section~\ref{pristine}).
That the odd antiferromagnetic state is the ground state agrees with Refs.~\cite{yoon17,koshino17} for the $(2,2)$ fault which is the same system as $N=4$ Bernal-stacked multilayer. For benchmarking, with $\alpha_g = 0.1$ we find $E_g = 1.42\,$meV for the $(2,2)$ fault which compares with $E_g \approx 1.44\,$meV for the second red circle in Fig.~3(a) of Ref.~\cite{yoon17}.

\begin{figure}[t]
\includegraphics[scale=0.2]{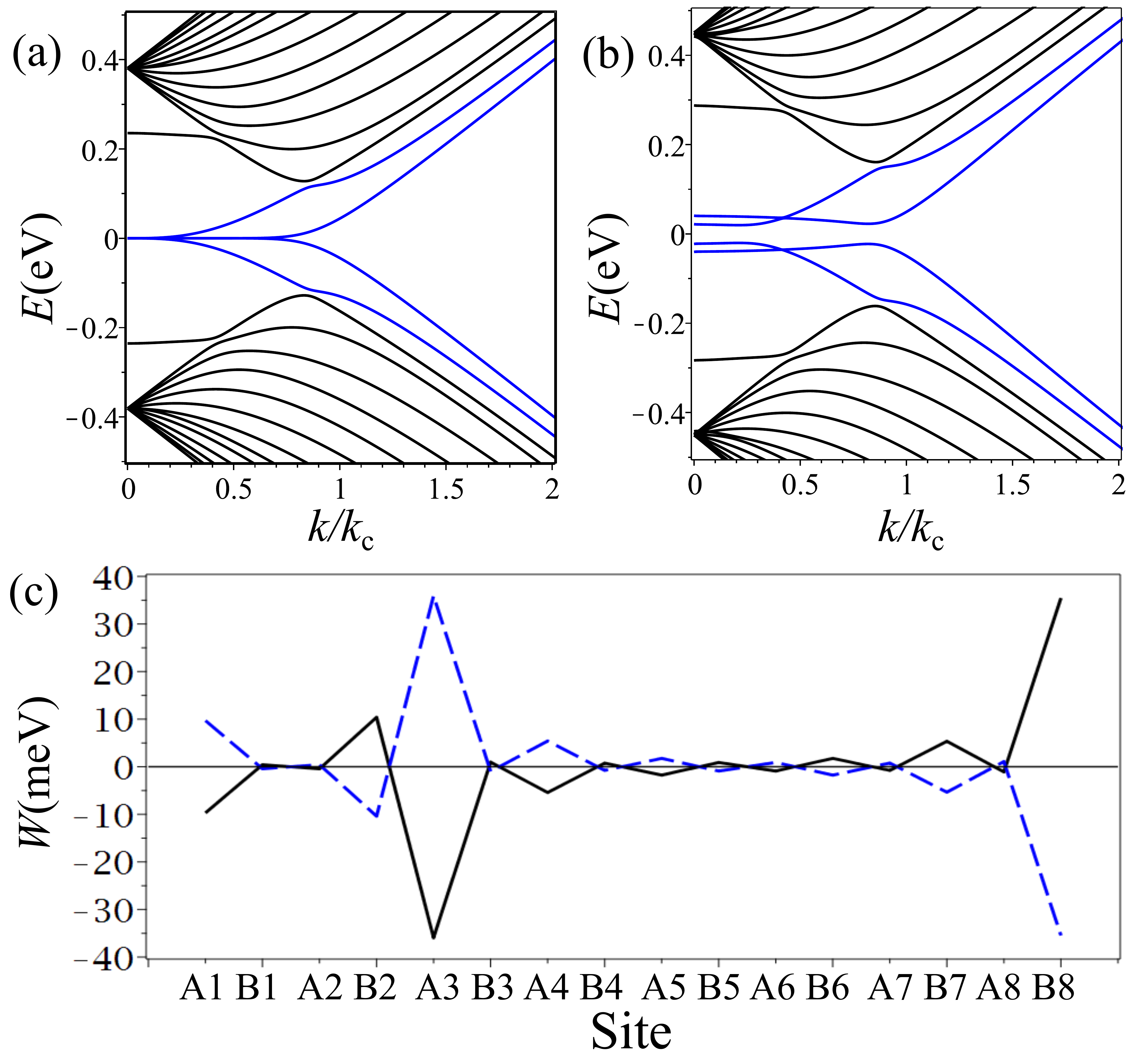}
\caption{Low-energy band structure of RMG with $N=16$ layers and a Bernal stacking fault between its third and fourth layers, $(m,n)=(3,13)$, with (a) non-interacting electrons described by Hamiltonian~(\ref{HNfull}) and (b) interacting electrons described by the mean-field theory Eqs.~(\ref{H0},\ref{VMF}) with interaction strength $\alpha_{\mathrm{g}} = 0.3$ and zero temperature. Black lines are bulk bands, blue lines are surface bands for $k \alt k_{\mathrm{c}}$ where $\hbar vk_{\mathrm{c}} = \gamma_1$.
(c) The values of the exchange potential $W_{k=0 \sigma X \!X}$ at the valley center $k=0$ and for each site $X^{\prime} = X$ for RMG with $N=8$ layers and a Bernal stacking fault off center~\cite{exchangecomment} between the second and third layers, $(m,n)=(2,6)$. The black solid line shows the exchange for two flavors, the blue dashed line is exchange for the other two flavors.
}\label{abfault-bs2}
\end{figure}

Since there are now four low-energy bands, we identify the separation at $k=0$ of the lowest conduction band and the highest valence band as $\Delta_1$ and the separation at $k=0$ of the second lowest conduction band and the second highest valence band as $\Delta_2$, $\Delta_2 \geq \Delta_1$.
For $m,n \gg 1$, we find that the low-energy bands of the interacting system behave as if they arise from two disconnected pieces of RMG.
This is not surprising as, due to their chirality, the wave functions of the $B_m$ and $A_{m+1}$ sites have different dependences on the polar angle in the graphene plane and, at $k=0$, this suppresses the exchange interaction matrix element between them.

In the special case of an even number of layers $N = 2m$ with a stacking fault on the central layer $(m,m)$, the low-energy bands are almost doubly degenerate (i.e. $8$-fold degenerate taking into account spin and valley), as shown for $N=16$ in Fig.~\ref{abfault-bs1}(a,b) for non-interacting and interacting electrons, respectively.
The exchange potential $W_{k=0 \sigma X \!X}$ at the valley center $k=0$ and for each site $X^{\prime} = X$ for RMG with $N=8$ layers and a Bernal fault at its center is shown in Fig.~\ref{abfault-bs1}(c)~\cite{exchangecomment}. This is a generalization of the odd antiferromagnetic state in pristine RMG, but now the exchange (and carrier density) per flavor has a substantial magnitude on sites $B_m$, $A_{m+1}$ by the stacking fault as well as the surface states.
The band gap, $E_{\mathrm{g}}$, and the order parameter $\Delta_1$ are plotted as data points in Fig.~\ref{abfault-gap1} as a function of layer number $N$ for different interaction strengths $\alpha_{\mathrm{g}}$ (for clarity, we don't plot $\Delta_2$ because $\Delta_2 \approx \Delta_1$ when the fault is at the center).
To illustrate that the system behaves almost as two separate RMG sections of $m = N/2$ layers,
the solid lines in Fig.~\ref{abfault-gap1} are not fits, but data taken from Fig.~\ref{abc-gap1} for a single RMG section with $N/2$ layers; for $N > 4$, the agreement is very close.

\begin{figure}[t]
\includegraphics[scale=0.21]{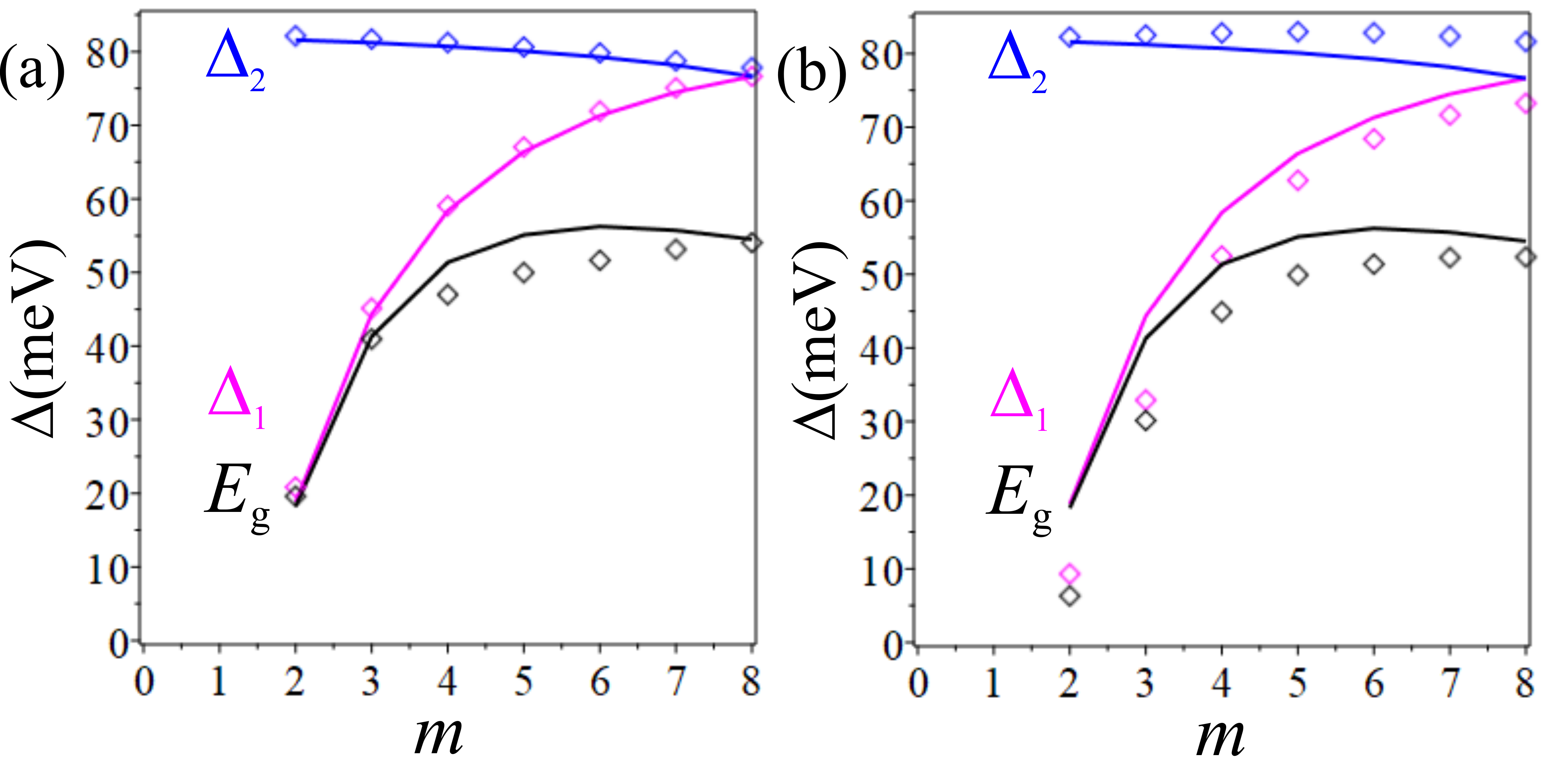}
\caption{(a) The band gap $E_{\mathrm{g}}$ and order parameters $\Delta_1$, $\Delta_2$ of RMG with $N=16$ layers and a Bernal stacking fault of structure $(m,16-m)$ consisting of a rhombohedral section of length $m$ coupled to a rhombohedral section of length $16-m$. Points are data for the system with the stacking fault, solid lines are data from Fig.~\ref{abc-gap1} for a single RMG section with $m$ layers for $\Delta_1$ and $E_{\mathrm{g}}$, and $16-m$ layers for $\Delta_2$. (b) The band gap $E_{\mathrm{g}}$ and order parameters $\Delta_1$, $\Delta_2$ of RMG with $N=15$ layers and a twin boundary fault of structure $(m,16-m)$ consisting of a rhombohedral section of length $m$ coupled to a rhombohedral section of length $16-m$. Points are data for the system with the stacking fault, solid lines are data from Fig.~\ref{abc-gap1} for a single RMG section with $m$ layers for $\Delta_1$ and $E_{\mathrm{g}}$, and $16-m$ layers for $\Delta_2$. For both plots, the interaction strength is $\alpha_{\mathrm{g}} = 0.3$ and temperature $T=0\,$K.
}\label{jplot1}
\end{figure}

More generally, the stacking fault breaks spatial inversion symmetry. 
As an example, bands for a $N = 16$ layer system $(m,n) = (3,13)$ with a three-layer section connected to a $13$-layer section are shown in Fig.~\ref{abfault-bs2}(a,b) for non-interacting and interacting electrons, respectively.
Although spatial inversion symmetry is absent, the band structure has flavor degeneracy for the antiferromagnetic ground state. This is not generally the case, {\it e.g.} within a ferrimagnetic configuration, the flavors with different orientation are usually not degenerate. For the interacting case, Fig.~\ref{abfault-bs2}(b) shows the bands of a single flavor for the antiferromagnetic ground state; within a flavor, the four low-energy bands are not degenerate.
In Fig.~\ref{abfault-bs2}(b), $\Delta_2 > \Delta_1$, where $\Delta_1$ (the separation at $k=0$ of the lowest conduction band and the highest valence band) is the order parameter related to the short section $m=3$, $\Delta_2$ (the separation at $k=0$ of the second lowest conduction band and the second highest valence band)
is related to the long section $n=13$.

The exchange potential $W_{k=0 \sigma X \!X}$ at the valley center $k=0$ and for each site $X^{\prime} = X$ for RMG with $N=8$ layers and a Bernal fault off center, $(m,n)=(2,6)$, is shown in Fig.~\ref{abfault-bs2}(c)~\cite{exchangecomment}. Again, this is a generalization of the odd antiferromagnetic state in pristine RMG. Although the off center stacking fault breaks spatial inversion symmetry, we refer to this as an odd parity state because the exchange has relative signs of (-,+,-,+) on the four low-energy orbitals ($A_1$, $B_2$, $A_{3}$, $B_8$). The exchange potential (and carrier density) per flavor has a larger magnitude on sites with low-energy orbitals $A_1$, $B_2$, $A_3$, $B_8$ sites, but has a much larger magnitude on sites  $A_3$, $B_8$ associated with the longer RMG section than $A_1$, $B_2$ related to the small section. This is reflected in the relative magnitudes of $\Delta_2$ and $\Delta_1$. 

\begin{figure}[t]
\includegraphics[scale=0.2]{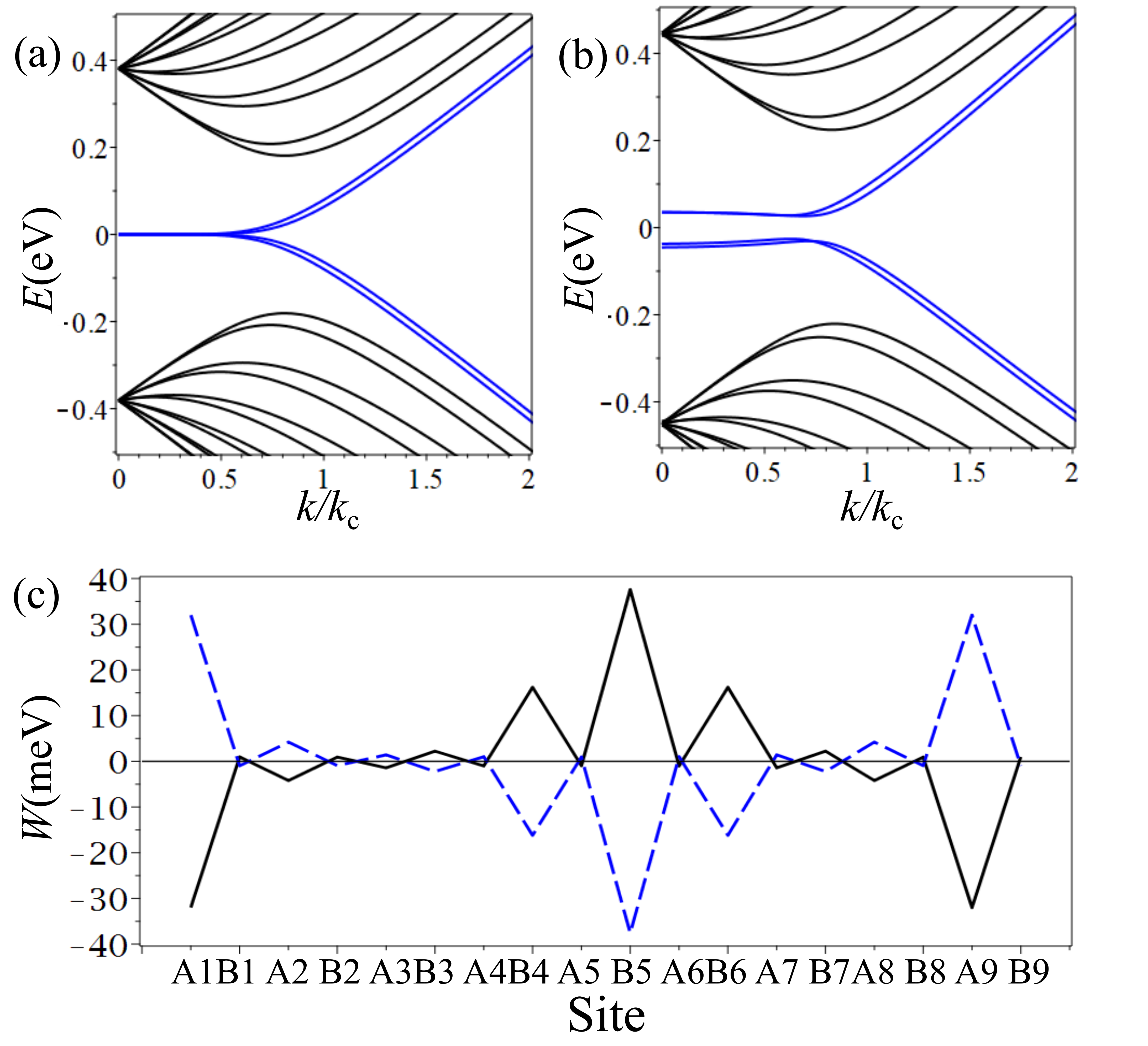}
\caption{Low-energy band structure of RMG with $N=15$ layers and a twin boundary stacking fault at its center with (a) non-interacting electrons described by Hamiltonian~(\ref{HNfull}) and (b) interacting electrons described by the mean-field theory~(\ref{H0},\ref{VMF}) with interaction strength $\alpha_{\mathrm{g}} = 0.3$ and zero temperature. Black lines are bulk bands, blue lines are doubly-degenerate low-energy bands for $k \alt k_{\mathrm{c}}$ where $\hbar vk_{\mathrm{c}} = \gamma_1$ ($8$-fold degenerate taking into account spin and valley degrees of freedom).
(c) The values of the exchange potential $W_{k=0 \sigma X \!X}$ at the valley center $k=0$ and for each site $X^{\prime} = X$ for RMG with $N=9$ layers and a twin boundary stacking fault at its center~\cite{exchangecomment}. The black solid line shows the exchange for two flavors, the blue dashed line is exchange for the other two flavors.
}\label{twin-bs1}
\end{figure}

Figure~\ref{jplot1}(a) shows $E_{\mathrm{g}}$, $\Delta_1$ and $\Delta_2$ for a $N=16$ system $(m,16-m)$ plotted as data points as a function of the number of layers $m$ in the short section ({\it i.e.} for different positions of the stacking fault). The solid lines are not fits, but they show data taken from Fig.~\ref{abc-gap1} for a single RMG section: a section of length $m$ is used to compare $E_{\mathrm{g}}$ and $\Delta_1$, a section of length $16-m$ is used to compare $\Delta_2$. For $\Delta_1$ and $\Delta_2$ the agreement is close, it is slightly less close for the band gap $E_{\mathrm{g}}$; $\Delta_1$ and $\Delta_2$ are determined at $k=0$ where the approximate splitting of the system into two parts is clearcut whereas $E_{\mathrm{g}}$ is generally determined at non-zero $k$. Weak coupling due to the Bernal stacking fault is indicated by the close agreement of the data points and lines for $\Delta_1$ and $\Delta_2$, as well as the nearly equal values of $\Delta_1$ and $\Delta_2$ for the spatially symmetric case $m=8$ ($\Delta_1$ and $\Delta_2$ differ by about $1\,$meV); the stacking fault only breaks the degeneracy slightly.

For the odd parity ground state, the Berry curvature $\Omega (k)$ is plotted in Fig.~\ref{berry1}(b) for $N=16$
with a stacking fault at the center $(m,n) = (8,8)$. This plot can be interpreted as the superposition of contributions from two identical sections of RMG summing to give a single peak. By contrast, Fig.~\ref{berry1}(c) shows $\Omega (k)$ for the odd parity ground state with the stacking fault off-center $(m,n) = (12,4)$. In this case, there are two separate peaks corresponding to the sections of length $m=12$ and $n=4$. For both of these cases, the Berry curvature $\Omega (k)$ sums to give a Chern number of magnitude $N/2$~\cite{fukui05,asboth16,chernvalley}.
As a further interesting example, Fig.~\ref{berry1}(d) shows $\Omega (k)$ for the same system in an even parity state [the exchange has relative signs of (-,+,+,-) on the four low-energy orbitals ($A_1$, $B_m$, $A_{m+1}$, $B_N$)] which is not the ground state. In this case, the contributions of the two sections of length $m=12$ and $n=4$ appear with an opposite sign and the Berry curvature $\Omega (k)$ sums to give a Chern number of magnitude $|m-n|/2$, {\it i.e.} it depends on the position of the stacking fault.

\section{Twin boundary stacking fault in rhombohedral graphene}\label{twin}

\subsection{Band structure of non-interacting electrons}

As a second example, we consider RMG with a twin boundary stacking fault, as illustrated in Fig.~\ref{lattice} (right panel) for $N = 7$ layers. Specifically, we use integers $(m,n)$ to denote
a rhombohedral section with $m$ layers and sites $A_1$, $B_1$, \ldots , $A_m$, $B_m$ connected by a twin boundary stacking fault to a rhombohedral section with $n$ layers and sites $B_{m}$, $A_{m}$, \ldots $B_N$, $A_N$ where $m \geq 2$, $n \geq 2$. The total layer number is $N = m+n-1$ because the two rhombohedral sections `share' the layer with sites $A_m$, $B_m$.
Thus, the example in Fig.~\ref{lattice} is a $(3,5)$ twin boundary stacking fault with $N=7$ total layers.
This fault contrasts with the Bernal fault. For example, at the stacking fault there are only three vertically connected atomic sites (sites $B_2$, $A_3$, $B_4$, in Fig.~\ref{lattice}) which make up a trimer; this is odd so it contributes two bulk bands and one band near zero energy at $k=0$~\cite{shi20} in a similar way to trilayer Bernal graphene~\cite{koshino09b}. Thus, overall, there are four low energy bands per spin and valley related to three sites not directly connected to a neighbor in an adjacent layer, namely $A_1$, $B_m$, $A_{N}$ (sites $A_1$, $B_3$, $A_7$ in Fig.~\ref{lattice}) plus
an odd combination of trimer sites $(B_{m-1}-B_{m+1})/\sqrt{2}$ [$(B_{2}-B_{4})/\sqrt{2}$ in Fig.~\ref{lattice}].

Since the low-energy orbitals near the stacking fault [i.e. those related to site $B_m$ and to $(B_{m-1}-B_{m+1})/\sqrt{2}$] are effectively shared between both sections of RMG either side of the fault, the two sides are more strongly coupled than in the Bernal fault case. For non-interacting electrons, the effective low-energy four band Hamiltonian, in a basis of orbitals on $A_1$, $(B_{m-1}-B_{m+1})/\sqrt{2}$, $A_{N}$, $B_m$ sites, is given by
\begin{widetext}
\begin{eqnarray}
H_{\mathrm{twin}}^{(m,n)} = \gamma_1\left(
           \begin{array}{cccc}
             0 & - (-\kappa^{\dagger})^{m-1}/\sqrt{2} & 0 & -(-\kappa^{\dagger})^{m}/2 \\
             - (-\kappa)^{m-1}/\sqrt{2} & 0 & (-\kappa)^{n-1}/\sqrt{2} & 0 \\
             0 & (-\kappa^{\dagger})^{n-1}/\sqrt{2} & 0 & -(-\kappa^{\dagger})^{n}/2 \\
             -(-\kappa)^{m}/2 & 0 & -(-\kappa)^{n}/2 & 0 \\
           \end{array}
         \right) . \label{htwin}
\end{eqnarray}
\end{widetext}
As the Hamiltonian is chiral, every matrix element between two A sites or between two B sites is zero. For the non-zero elements (between A and B sites), we keep only the leading terms in $k/k_c$.
The second and fourth columns indicate that the B orbitals at the fault are coupled to both of the RMG sections.

Since the four band Hamiltonian~(\ref{htwin}) is chiral, the energy spectrum of non-interacting electrons displays electron-hole symmetry and the band energies $E$ are given by
\begin{eqnarray*}
( E /\gamma_1 )^2 &=& \frac{1}{2} \beta_{\mathrm{t}} (k/k_c) \pm \frac{1}{2} \sqrt{\beta_{\mathrm{t}}^2 (k/k_c) - 4 \eta_{\mathrm{t}} (k/k_c)} , \\
\beta_{\mathrm{t}} (x) &=& \frac{1}{2} \! \left( x^{2m-2} + x^{2n-2} \right) +  \frac{1}{4} \! \left( x^{2m} + x^{2n} \right) , \\
\eta_{\mathrm{t}} (x) &=&  \frac{1}{2}x^{2m+2n-2} .
\end{eqnarray*}
When the fault lies exactly in the center of a long RMG system: $n=m = (N+1)/2 \approx N/2$ with $N \gg 1$,
then $E \approx \pm \gamma_1 (k/k_c)^{N/2}$ and $E \approx \pm (\gamma_1 / \sqrt{2}) (k/k_c)^{N/2}$. The low-energy dispersion of a pair of the bands acquires an additional factor of $1/\sqrt{2}$ as compared to the dispersion of a stack with $N/2$ layers, indicating that the twin boundary stacking fault strongly affects the electronic behavior of the system.

\subsection{Numerical mean-field theory}

The numerical mean-field theory calculations proceed as for pristine RMG, except that the stacking fault is taken into account by a different position of the interlayer coupling $\gamma_1$ in the non-interacting Hamiltonian~(\ref{HNfull}).
In the special case of an odd number of layers $N = 2m-1$ with a stacking fault on the central layer $(m,m)$, the low-energy bands are almost doubly degenerate (i.e. $8$-fold degenerate taking into account spin and valley), as shown for $N=15$ in Fig.~\ref{twin-bs1}(a,b) for non-interacting and interacting electrons, respectively.
Although it is not clearly visible in Fig.~\ref{twin-bs1}(b), the low-energy bands actually have a small separation of a few meV (this is indicated in Fig.~\ref{jplot1}(b) where there is a non-zero separation of $\Delta_1$ and $\Delta_2$ for $m=8$).

\begin{figure}[t]
\includegraphics[scale=0.19]{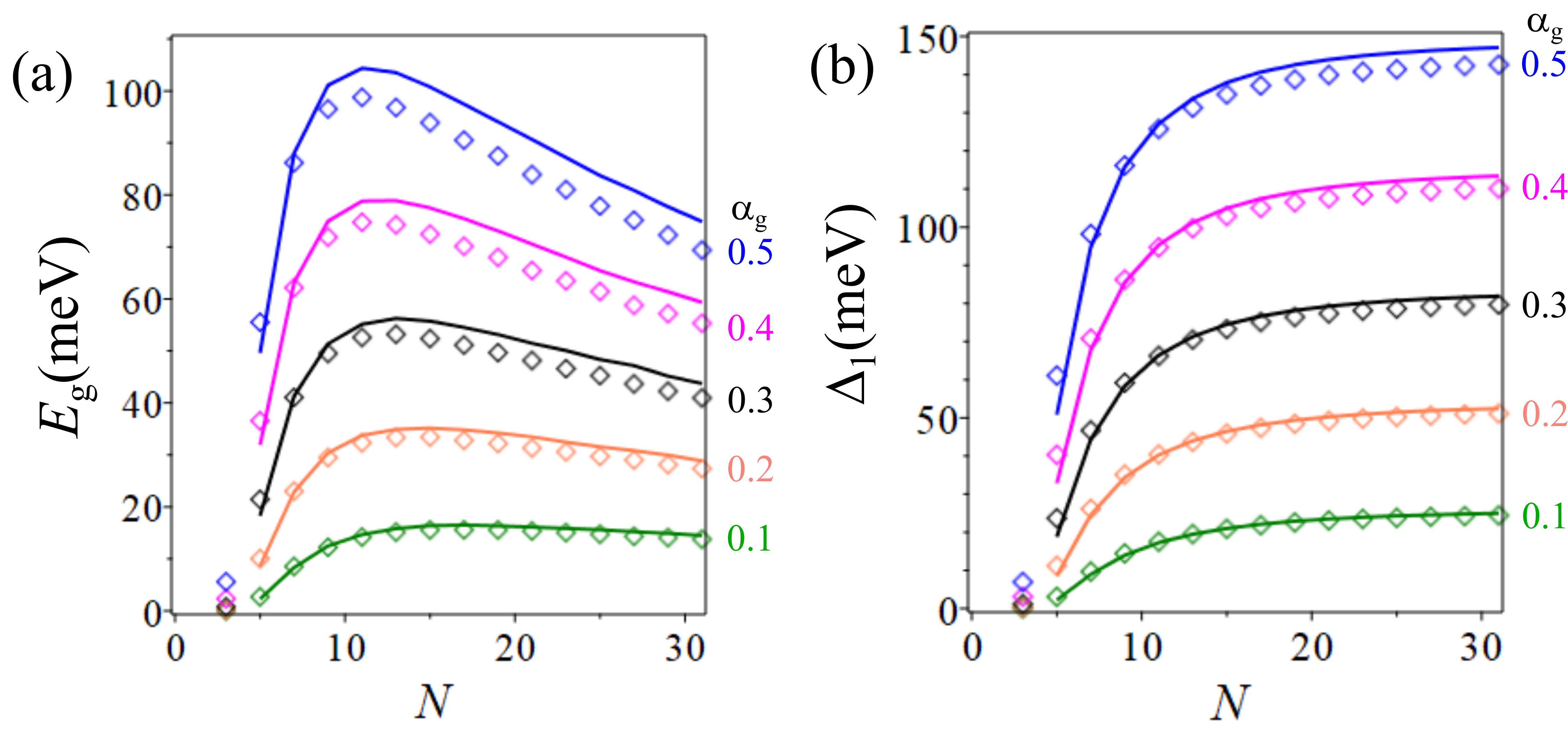}
\caption{The band gap of RMG with a twin boundary stacking fault at its center, for odd $N$, at zero temperature as a function of layer number $N$ for different values of interaction strength $\alpha_{\mathrm{g}}$ showing (a)  the band gap $E_{\mathrm{g}}$ and (b) the order parameter $\Delta_1$. Points are data for the system with the stacking fault, solid lines are data from Fig.~\ref{abc-gap1} for a single RMG section with $(N-1)/2$ layers.
}\label{twin-gap1}
\end{figure}

The exchange potential $W_{k=0 \sigma X \!X}$ at the valley center $k=0$ and for each site $X^{\prime} = X$ for RMG with $N=9$ layers and a fault at its center is shown in Fig.~\ref{twin-bs1}(c)~\cite{exchangecomment}.
This ground state has an even parity of exchange (and carrier density) per flavor, with the large magnitude of exchange on the surface orbitals $A_1$, $A_9$ having the same sign, low-energy orbitals at the fault $B_4$, $B_5$, $B_6$ have exchange potentials with the opposite sign. Within the four spin-valley flavors, the ground state has an antiferromagnetic configuration, because this minimizes the Hartree energy as previously. 

The band gap, $E_{\mathrm{g}}$, and the order parameter $\Delta_1$ are plotted as data points in Fig.~\ref{twin-gap1} as a function of layer number $N$ for different interaction strengths $\alpha_{\mathrm{g}}$ and a fault at the center (for clarity, we don't plot $\Delta_2$ because $\Delta_2 \approx \Delta_1$ when the fault is at the center).
The solid lines are data taken from Fig.~\ref{abc-gap1} for a single RMG section with $(N-1)/2$ layers; choosing $(N-1)/2$ gives generally better agreement of solid lines and data points than choosing $N/2$ (as in Fig.~\ref{abfault-gap1}).
There is good agreement of the solid lines and data points, but not as close as for the Bernal fault, Fig.~\ref{abfault-gap1}.

\begin{figure}[t]
\includegraphics[scale=0.2]{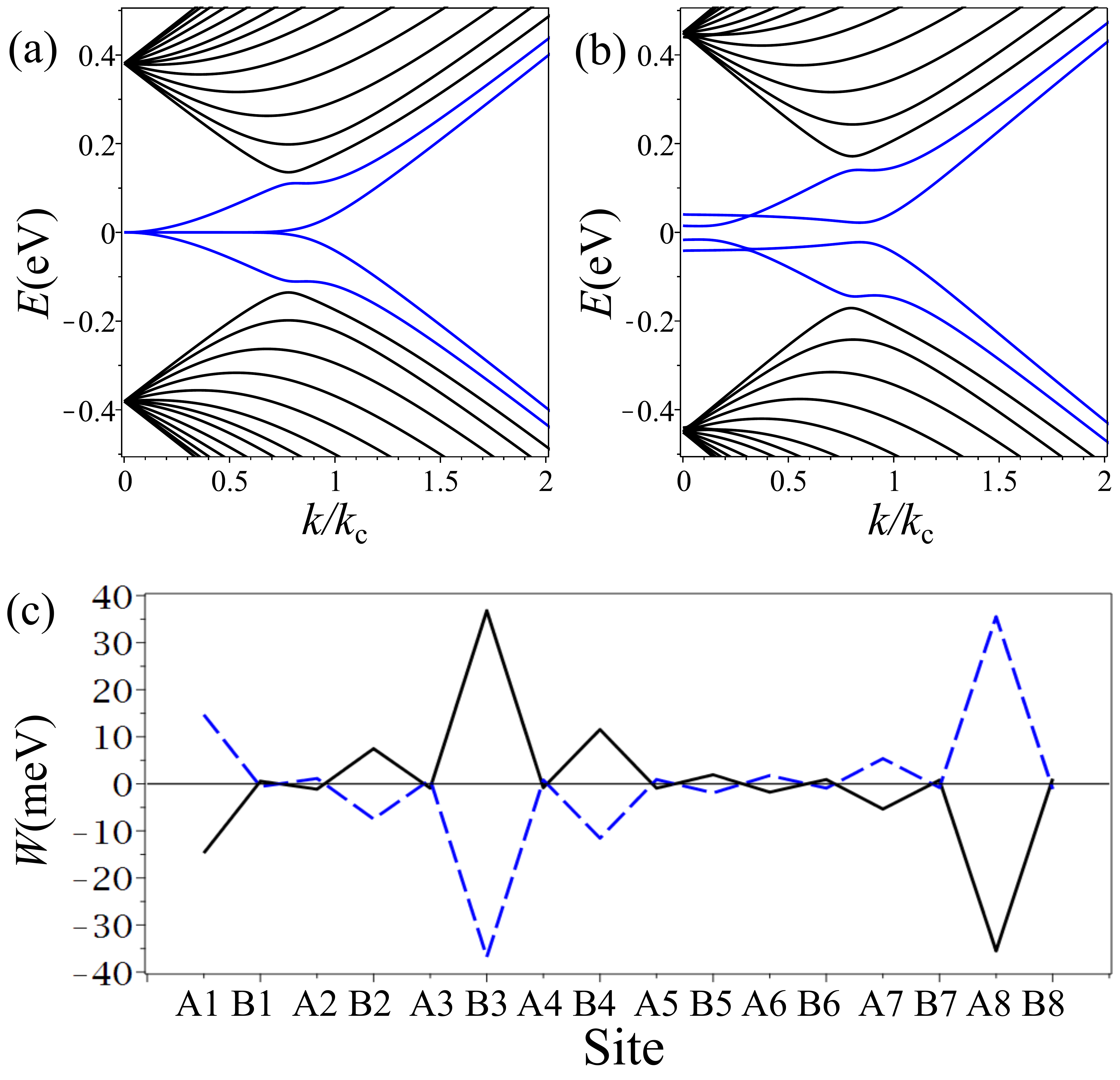}
\caption{Low-energy band structure of RMG with $N=16$ layers and a twin boundary fault off center at its third layer, $(m,n)=(3,14)$, with (a) non-interacting electrons described by Hamiltonian~(\ref{HNfull}) and (b) interacting electrons described by the mean-field theory Eqs.~(\ref{H0},\ref{VMF}) with interaction strength $\alpha_{\mathrm{g}} = 0.3$ and zero temperature. Black lines are bulk bands, blue lines are surface bands for $k \alt k_{\mathrm{c}}$ where $\hbar vk_{\mathrm{c}} = \gamma_1$.
(c) The values of the exchange potential $W_{k=0 \sigma X \!X}$ at the valley center $k=0$ and for each site $X^{\prime} = X$ for RMG with $N=8$ layers and a twin boundary stacking fault off center~\cite{exchangecomment} at its third layer, $(m,n)=(3,6)$. The black solid line shows the exchange for two flavors, the blue dashed line is exchange for the other two flavors.
}\label{twin-bs2}
\end{figure}

Bands for a $N = 16$ layer system $(3,14)$ with an off center fault, namely a three-layer section connected to a $14$-layer section are shown in Fig.~\ref{twin-bs2}(a,b) for non-interacting and interacting electrons, respectively.
The exchange potential $W_{k=0 \sigma X \!X}$ at the valley center $k=0$ and for each site $X^{\prime} = X$ for RMG with $N=8$ layers and a fault off center is shown in Fig.~\ref{twin-bs2}(c)~\cite{exchangecomment}. This is also the even parity state [the exchange has relative signs of (-,+,+,+,-) on the low-energy orbitals ($A_1$, $B_{m-1}$, $B_{m}$, $B_{m+1}$, $A_N$)].
In the interacting case, Fig.~\ref{twin-bs2}(b), the four low-energy bands are clearly not degenerate.
Figure~\ref{jplot1}(b) shows $E_{\mathrm{g}}$, $\Delta_1$ and $\Delta_2$ for a $N=15$ system $(m,16-m)$ plotted as data points as a function of the number of layers $m$ in the short section. The solid lines show data taken from Fig.~\ref{abc-gap1} for a single RMG section: a section of length $m$ is used to compare $E_{\mathrm{g}}$ and $\Delta_1$, a section of length $16-m$ is used to compare $\Delta_2$.
The agreement of the data points and the solid lines is reasonable, although not as close as in the case of the Bernal fault Fig.~\ref{jplot1}(a), this is due to the larger coupling between the two RMG sections in the twin stacking fault case. In particular, when the stacking fault is in the center and doesn't break spatial inversion symmetry, $m=8$, there's still a significant difference between $\Delta_1$ and $\Delta_2$ (of about $8\,$meV), whereas this difference is small (about $1\,$meV) in the Bernal fault case, Fig.~\ref{jplot1}(a).

\begin{figure}[t]
\includegraphics[scale=0.2]{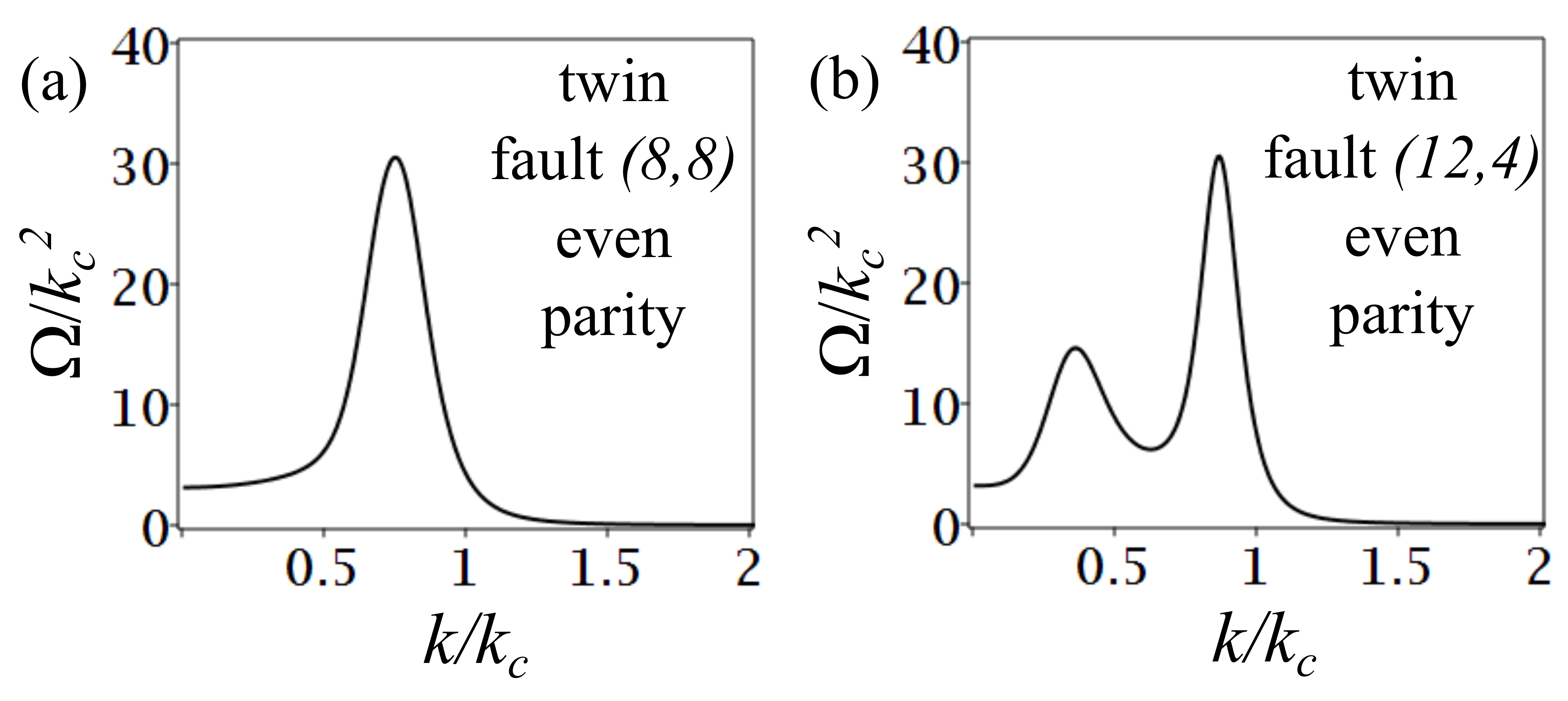}
\caption{Berry curvature $\Omega (k)$ as a function of the magnitude of the wave vector $k$ plotted from the valley center (the Berry curvature is isotropic in the minimal model) for $N=15$, $\alpha_{\mathrm{g}} = 0.3$ and $T=0\,$K. (a) is for a twin boundary fault at the center in the even parity ground state, (b) is for a twin boundary fault off center in the even parity ground state.
}\label{berry2}
\end{figure}

For the even parity ground state, the Berry curvature $\Omega (k)$ is plotted in Fig.~\ref{berry2}(a) for $N=15$
with a stacking fault at the center $(m,n) = (8,8)$. This plot can be interpreted as the superposition of contributions from two identical sections of RMG summing to give a single peak. By contrast, Fig.~\ref{berry2}(b) shows $\Omega (k)$ for the even parity ground state with the stacking fault off-center $(m,n) = (12,4)$. In this case, there are two separate peaks corresponding to the sections of length $m=12$ and $n=4$. For both of these cases, the Berry curvature $\Omega (k)$ sums to give a Chern number of magnitude $N/2$~\cite{fukui05,asboth16,chernvalley}.
Although these are even parity states, they give the same Chern numbers as the odd parity states in the system with a Bernal fault. The reason is that the twin boundary fault effectively flips the position of $A$ and $B$ sites within a layer for the layers above the fault (right panel of Fig.~\ref{lattice}) compensating the change of relative sign of potential differences for the even parity state.

\section{Temperature dependence of the order parameters}\label{temp}

\begin{figure}[t]
\includegraphics[scale=0.22]{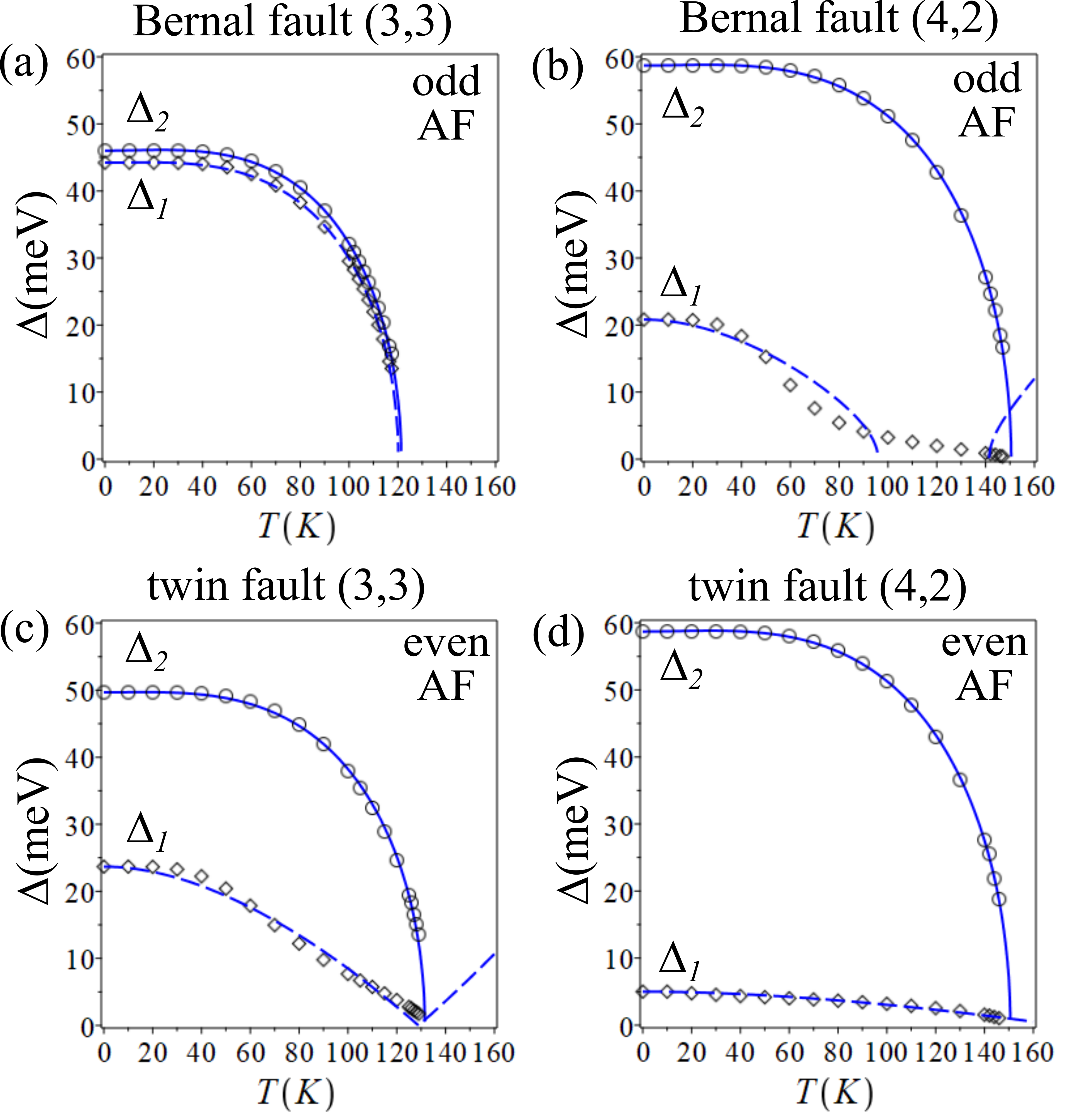}
\caption{Temperature dependence of the order parameters $\Delta_1$ and $\Delta_2$ for RMG with $N=6$ layers and the odd antiferromagnetic state for (a) a Bernal stacking fault at the center and (b) a Bernal stacking fault off center. The order parameters for $N=5$ and the even antiferromagnetic state for (c) a twin boundary stacking fault at the center and (d) a twin boundary stacking fault off center. Circles show data for $\Delta_2$, diamonds show data for $\Delta_1$, lines (solid for $\Delta_2$ and dashed for $\Delta_1$) show fits using Eq.~(\ref{tcfit}). For all plots, $\alpha_{\mathrm{g}} = 0.3$.
}\label{fig15}
\end{figure}

For an isolated stacking fault in RMG, we consider how the transition temperature for the mean field AF state is determined in the presence of two order parameters. To do this, we studied the simplest case of RMG with $N=6$ layers which can have either a Bernal stacking fault at the center, $(m,n) = (3,3)$, or off center, $(m,n) = (2,4)$. Finite temperature is taken into account through the Fermi-Dirac distribution in the mean field theory Eqs.~(\ref{hartree},\ref{exchange}), and the temperature dependence of the order parameters is shown in Fig.~\ref{fig15}(a) and (b). Flavor degeneracy is generally broken (when spatial inversion symmetry is absent), resulting in slightly different values of $\Delta_1$ and $\Delta_2$ for two flavors as compared to the other two: in the figures we plot the smallest values. The value of the band gap $E_g$ is very close to that of $\Delta_1$, $E_g \alt \Delta_1$ in general.

For the Bernal fault at the center, Fig.~\ref{fig15}(a), there is weak coupling between the two sections of RMG as indicated by the small difference between $\Delta_1$ and $\Delta_2$ at $T=0\,$K. The temperature dependence of $\Delta_1$ and $\Delta_2$ is very similar, and similar to the behavior of trilayer RMG Fig.\ref{abc-tc1}(a), and they converge to the same critical temperature $T_c \approx 120\,$K which is close to that of trilayer, $T_c = 118\,$K, Fig.\ref{abc-tc1}(b).

For a Bernal fault off center, Fig.~\ref{fig15}(b), spatial inversion symmetry is absent and the behavior is different. At $T=0\,$K, the values of $\Delta_1 \approx 20\,$meV and $\Delta_2 \approx 60\,$meV are similar to those of RMG with $N=2$ and $N=4$, respectively, Figs.~\ref{abc-gap1}(b) and~\ref{abc-tc1}(a).
The temperature dependence of $\Delta_2$ is described well by the fit~(\ref{tcfit}) and $\Delta_2$ behaves in a similar way to the order parameter of $N=4$ RMG, Fig.~\ref{abc-tc1}(a). However, $\Delta_1$ doesn't follow the behavior of $N=2$ RMG, Fig.~\ref{abc-tc1}(a), which has $T_c \approx 50\,$K, and the fit of $\Delta_1$ to Eq.~(\ref{tcfit}) is very poor.
Despite the apparently weak coupling between the two RMG sections at $T=0\,$K, once the magnitude of $\Delta_1$ falls at finite temperature, its behavior is strongly influenced by the larger section. Thus $\Delta_1$ (and the band gap $E_g \alt \Delta_1$) remain non-zero all the way up to $T_c \approx 150\,$K determined by $\Delta_2$ (which is close to that of $N=4$ RMG, $T_c = 151\,$K, Fig.~\ref{abc-tc1}).
Overall, this means that when a Bernal stacking fault is off center separating the system into a long and short section, the band gap at zero temperature is determined by the short section ($\Delta_1$), but the transition temperature $T_c$ is determined by the long section ($\Delta_2$).

Fig.~\ref{fig15}(c) shows the temperature dependence of the AF order parameters for a twin boundary fault at the center of an $N=5$ system, Fig.~\ref{fig15}(d) shows an off center twin fault. For the symmetric system, Fig.~\ref{fig15}(c), $\Delta_2$ and $\Delta_1$ are quite different at $T=0\,$K because of stronger coupling by the twin fault than the Bernal one. For both cases, Fig.~\ref{fig15}(c) and (d), the temperature dependence of $\Delta_2$ fits Eq.~(\ref{tcfit}) very well and $T_c$ ($\approx 131\,$K and $\approx 151\,$K, respectively) is close to that of a RMG system with $N=3$ and $N=4$ layers, respectively; the $\Delta_2$ plots are similar to those of a Bernal fault in panels (a) and (b). Fits to $\Delta_1$ are not as good, and the value of fitting parameter $A$ in the fit Eq.~(\ref{tcfit}) is far smaller than for $\Delta_2$ or for pristine RMG, indicating that $\Delta_1 (T)$ is quite different here. As for the off center Bernal fault, $\Delta_1$ (and the band gap $E_g$) remain non-zero [albeit of small magnitude in Fig.~\ref{fig15} (d)] up to the $T_c$ determined by $\Delta_2$.

\section{Discussion}

We have generalized the mean-field Hartree Fock description~\cite{min08,jung11,jung13,yoon17,koshino17} to provide a comprehensive qualitative description of broken symmetry ground states in RMG, including the effects of defects including random disorder and stacking faults.
The non-trivial topology of the low-energy bands is reflected in large Berry curvature and Chern numbers per spin-valley flavor.
An obvious generalization is to a number of stacking faults separating rhombohedral sections with different numbers of layers $m$, each section contributing a pair of low-energy flat bands and a peak in the Berry curvature at a characteristic $m$-dependent wave vector.
In RMG, each stacking fault contributes a pair of low-energy flat bands because they are more complicated than the domain walls usually considered in the SSH model which consist of isolated monomers or trimers~\cite{asboth16}. The Bernal fault corresponds to two monomers (and a $4$-mer), the twin boundary fault is a monomer plus a trimer.

As described in Section~\ref{fbm}, sources of systematic uncertainty include the choice of cutoff $k_{\star}$, the omission of tight-binding parameters in the minimal model, and the value of the interaction parameter $\alpha_{\mathrm{g}}$.
We have assumed the interaction parameter $\alpha_{\mathrm{g}}$ to be independent of layer number $N$, but it is anticipated that the effective strength of interactions could fall with $N$ due to screening~\cite{wehling11,jia13}. This would lead to a further reduction in $E_{\mathrm{g}}$ and a fall in $\Delta$ for large $N$ in Fig.~\ref{abc-gap1}, say.
Additional tight-binding parameters such as $\gamma_2$ and $\gamma_3$ will introduce trigonal warping of the dispersion around each valley (so the Berry curvature, Fig.~\ref{berry1}, will be anisotropic), and $\gamma_4$ will break particle-hole symmetry~\cite{koshino09,slizovskiy19}; this is likely to reduce the value of the band gap. The additional tight-binding parameters are usually smaller in magnitude than the typical values of the band gap that we predict, but there is a possibility that additional parameters will change the qualitative nature of the ground state~\cite{koshino17}.  However, even without these parameters, for our choice of cutoff and for $\alpha_{\mathrm{g}} = 0.3$, we find close agreement of band gap values in RMG with DFT calculations of Ref.~\cite{pamuk17} (which considered $N=3$ to $N=8$ layers).

The mean-field Hartree Fock approach neglects strong correlation effects, and there have been predictions of magnetic ordering~\cite{otani10,xu12} and superconductivity~\cite{kopnin13,munoz13,lothman17} due to the flat bands in RMG. We speculate that the additional flat bands localized at stacking faults, and in close spatial proximity to each other, are more likely to support strongly-correlated states than the widely-separated surface states in pristine RMG.

All relevant data present in this publication can be accessed at \cite{datalink}.

\acknowledgments

The authors thank Mikito Koshino for discussions.
Computer time was provided by the Lancaster University High End Computing facility.


\begin{thebibliography}{99}
\bibitem{chernvalley}
We calculate the Chern number per spin-valley flavor by integrating the non-Abelian Berry curvature~\cite{fukui05} for the occupied valence bands in the vicinity of a single valley. Since this is not the actual Chern number (obtained by integrating over the whole Brillouin zone), the Chern number per flavor may take half integer values.
\bibitem{fukuicomment}
Note that definitions of intermediate quantities Berry curvature and Berry connection differ by factors such as $i$ between~\cite{fukui05} and~\cite{asboth16}, although these sources agree in the relation between Chern number and wave functions.
\bibitem{exchangecomment}
When plotting the exchange interaction $W_{k=0 \sigma X \!X}$, we subtract a constant term so that the mean value is zero.
\end{thebibliography}
\end{document}